\documentstyle[aps,preprint]{revtex}
\input{psfig}
\tightenlines

\begin{document}

\title{Orthorhombic Phase of Crystalline Polyethylene: A Constant Pressure Path 
Integral Monte Carlo Study}

\author{R. Marto\v{n}\'ak$^{(1,2,}$\cite{byline}$^)$, W. Paul$^{(1)}$,
K. Binder$^{(1)}$}

\address{
$^{(1)}$ Institut f\"ur Physik, KoMa 331, Johannes Gutenberg--Universit\"at \\
Staudingerweg~7, 55099~Mainz, Germany \\
$^{(2)}$ Max--Planck--Institut f\"ur Polymerforschung \\
Ackermannweg~10, 55021~Mainz, Germany \\}

\date{\today}
\maketitle

\begin{abstract}
\noindent
In this paper we present a Path Integral Monte Carlo (PIMC) simulation of the 
orthorhombic phase of crystalline polyethylene, using an explicit atom force 
field with unconstrained bond lengths and angles.
This work represents a quantum extension of our recent classical simulation \cite{rm}.
It is aimed both at exploring the applicability of the PIMC method on such polymer 
crystal systems, as well as on a detailed assessment of the importance of quantum 
effects on different quantities. We used the $NpT$ ensemble and simulated the system 
at zero pressure in the temperature range 25 -- 300 K, using Trotter numbers 
between 12 and 144. In order to investigate finite-size effects, we used chains 
of two different lengths, C$_{12}$ and C$_{24}$, corresponding to the total
number of atoms in the super-cell being 432 and 864, respectively. We show here the 
results for structural parameters, like the orthorhombic lattice constants $a,b,c$, 
and also fluctuations of internal parameters of the chains, such as bond lengths and 
bond and torsional angles.
We have also determined the internal energy and diagonal elastic constants 
$c_{11}, c_{22}$
and $c_{33}$. We discuss the temperature dependence of the measured quantities and 
compare to that obtained from the classical simulation. For some quantities, we 
discuss the way they are related to the torsional angle fluctuation.
In case of the lattice parameters we compare our results to those obtained from other 
theoretical approaches as well as to some available experimental data. In order to
study isotope effects, we simulated also a deuterated polyethylene crystal at low
temperature. We also suggest possible ways of extending this study and present some
general considerations concerning modeling of polymer crystals.

\end{abstract}

\draft

\section{Introduction}

Polymer crystals are well known to be intrinsically difficult to prepare in a highly
crystalline state, which in turn hinders the possibilities of their experimental 
characterization.
As a consequence, computer simulation appears to be a convenient tool to study their
properties. For crystalline PE, which represents the simplest and thus paradigmatic 
case, it has been recently shown \cite{rm} that classical constant pressure MC is 
a well applicable simulation method, provided a good quality force field
is available. It allows to calculate the whole variety of static local and collective
quantities, including properties of major practical and technological importance,
like thermal expansion and elastic constants. On the other hand,
recent work on the same system, using quasi-harmonic or self-consistent quasi-harmonic 
approximation \cite{rutledge,hagele,stobbe}, has clearly pointed to a quantitative as
well as a qualitative inadequacy of the classical treatment at low temperatures, where 
quantum effects cannot be neglected anymore.

Generally, quantum effects are known to be
important in lattice dynamics of solids in the low-temperature region, when classical 
thermal fluctuations become comparable or smaller than the amplitude of the quantum 
zero-point motion. Under particular circumstances, when two different 
crystal structures have classically very close energies, and the system is close 
to a structural phase transition, quantum fluctuations can play a decisive role. 
In case of solid nitrogen \cite{scott}, they are responsible for a strong 
isotope effect on the low-temperature $\alpha - \gamma$ structural phase transition 
as a function of pressure. If the classical energy difference is small enough, 
quantum effects can even suppress the
transition altogether, and stabilize the disordered phase, as it happens in the case 
of quantum paraelectrics SrTiO$_3$ or KTaO$_3$, where the ferroelectric long-range 
order is only incipient down to $T = 0$ \cite{mb}. Even though a PE crystal does not
represent such a dramatic case, at low temperatures it is not possible to account 
even qualitatively for the temperature dependence of quantities like thermal expansion
coefficients and elastic constants without quantum effects being duly taken into
account.

A natural extension of the classical MC method in order to include quantum effects at 
finite temperature is the Path Integral Monte Carlo scheme \cite{ceperley}. Recently, 
the $NpT$ version of this
method has been applied to study quantum effects in crystals at low temperatures,
in particular in solid rare gas systems \cite{mueser}, and silicon crystal
\cite{rickward}. In case of polymer crystals, the distinguishing feature is an
extreme anisotropy, closely related to the existence of many energy scales, ranging 
from soft intermolecular (non-bonded) interactions to stiff intramolecular (bonded) 
interactions. This feature
presents a problem already at the classical level, as discussed in \cite{cpc,jcamd,rm},
and requires an introduction of special global moves in the sampling algorithm.
In the quantum case, we should moreover expect very different
convergence properties of different physical quantities as a function of Trotter 
number, depending on the typical energy scale with which a given quantity is 
associated.

The aim of the paper is basically twofold. On the one hand, we wanted to explore the
applicability of constant pressure PIMC method to a PE crystal, and determine the 
region of temperatures where the use of 
the method is practical. On the other hand, since the PIMC scheme is capable of providing 
essentially exact results, it can also be used to assess the range of validity 
of approximate analytical methods, like, e.g., quantum quasi-harmonic approximation 
\cite{rutledge,hagele,stobbe}. This is particularly important for the study of intrinsically
anharmonic phenomena, like, e.g., lattice thermal expansion.

The paper is organized as follows. In section \ref{pimc}, we briefly describe the PIMC
simulation method used, without addressing the force field and its implementation, since these 
issues have already been discussed in detail in \cite{rm}. In section \ref{resdis}, 
we present and discuss the results, paying particular attention to comparison
of the quantum results to the classical ones. In the final section \ref{conc} we draw
some conclusions and suggest some possible ways of extending this study. We also
make several remarks concerning general issues related to modeling of polymer 
crystals. For completeness, in Appendix A we present the full form of the force field we have
used together with numerical values of the parameters. In Appendix B, we present
some considerations on the relation of 
correlation functions of torsional fluctuations to the contraction of the crystal
along the chain axis.

\section{PIMC simulation method}
\label{pimc}

In this section, we will describe only those features of the simulation method which
are specific for the constant pressure PIMC scheme. The implementation of the
SLKB force field \cite{slkb} we used as well as 
many other features of the present algorithm are exactly identical to those
of our classical simulation, which has been described in detail in Ref.~\cite{rm}.
For convenience, however, in Appendix A we summarize the form and parameters of the 
force field.

We have implemented the constant pressure PIMC scheme basically along the same lines 
as it was done for a cubic system in Ref.~\cite{mueser}, the only difference being 
that in our case we had
to use an anisotropic version of the $NpT$ ensemble. We have used the primitive 
decomposition of the hamiltonian, resulting in the effective hamiltonian
\begin{equation}
H_P (\{ \vec{r}_{i}^{k} \}) = \sum_{k=1}^{P} \left( \sum_{i=1}^{N} 
{{m_i P}\over{2 \hbar^2 \beta^2}} 
\left( \vec{r}_{i}^{k} - \vec{r}_{i}^{k-1} \right)^2 + 
{{1}\over{P}} V(\{ \vec{r}_{i}^{k} \}) \right) \; , \label{efham}
\end{equation}
where $N$ is the number of particles in the quantum system, $m_i$ are their masses, 
$P$ is the Trotter number, $\beta = 1/k_B T$ is the inverse temperature and 
$V(\{ \vec{r}_{i} \})$ is the potential energy of the system. Such an effective hamiltonian 
represents a pseudo-classical system consisting of $P$ copies (Trotter slices) of the 
original system, individual particles in neighboring Trotter slices being connected 
via harmonic "springs", and periodic boundary conditions being applied along the 
Trotter direction. This pseudo-classical system has now $N P$ particles and can be 
simulated using the same constant pressure MC algorithm as in the classical case. The 
acceptance criterion
for the volume moves was based on the Boltzmann factor $(s_1 s_2 s_3)^{N P}
e^{-\beta E_P}$, where $E_P = H_P + p V_0 s_1 s_2 s_3$, $p$ is the external pressure, 
$V_0$ is the volume of the reference super-cell and $s_1,s_2,s_3$ are three independent
scaling factors along the coordinate axes. In all simulations described in this
paper, the external pressure was set to zero. The estimators for all configurational 
properties, diagonal in the coordinate representation, are straightforward analogues
of their classical counterparts, while for the kinetic energy we used the virial estimator
\cite{hbb}.

To sample the system, we have used three kinds of moves: classical moves, quantum moves
and volume moves. Classical moves of two types, local moves of atoms or global 
moves of whole polymer chains, as described in Ref.~\cite{cpc,jcamd,rm}, have now always 
been applied to all particles or chains with a given 
number in all Trotter slices simultaneously. These moves sample the classical 
configurational phase space of the system. We note here that when performing a rotation
of a given chain in all Trotter slices, the energies of the "harmonic springs" between
the corresponding individual particles have to be recalculated explicitly, in contrast
to pure translational moves of a particle in all Trotter slices, which preserve the 
energy of the "springs". Quantum moves consisted of local translational moves of 
individual particles of the pseudo-classical system, which sample the quantum 
fluctuations around the classical paths. In the quantum moves, different maximum 
displacements have been used for C and H atoms, not only because of the different 
number of bonds but also because of the different mass of the atoms and 
resulting different stiffness of the "springs". In volume moves,
we performed a simultaneous rescaling of coordinates of all particles in all Trotter 
slices by the three scaling factors $s_1,s_2,s_3$. One Monte Carlo step per 
site (MCS) thus 
consisted of an attempted quantum move on each particle of the pseudo-classical
system, followed by a classical move attempted successively on all atoms or all chains
(always simultaneously in all Trotter slices, as described above) and a volume move.
Among the classical moves, 30 \% of global moves were used, like in the classical study 
\cite{rm}. For all kinds of moves, the displacements were chosen to yield an 
acceptance ratio of 20 - 30 \%.

We have simulated systems with C$_{12}$ and C$_{24}$ chains, consisting of 
$2 \times 3 \times 6$ and $2 \times 3 \times 12$ unit cells, respectively (432 and 864 
atoms), at temperatures 25, 50, 100, 150, 200 and 300 K. At different temperatures, different
numbers of values of the Trotter number were used. For the smaller system, we used at 
25 K $P = 144$, at 50 K $P = 54, 72, 144$, at 100 K $P = 36, 54, 72$, at 150 K $P = 48$,
at 200 K $P = 16, 24, 32$ and finally at 300 K $P = 12, 16, 24$. The larger system was
simulated only at 100 K with $P = 72$ and at 300 K with $P = 24$. We note here that the
largest pseudo-classical systems simulated consisted of $432 \times 144 = 864 \times 72
= 62208$ 
particles, which is quite a large number. As an initial configuration for a given
temperature, we always used a pseudo-classical system consisting of $P$ identical copies
of an equilibrated classical configuration at the same temperature. This configuration
was then equilibrated for several thousands MCS with the full PIMC algorithm, which 
corresponded to switching on the quantum fluctuations and allowing the system to find 
a new equilibrium. For illustration of the run length used for measurement, for the 
smaller system at 300 K and $P = 24$ we used about 280000 MCS. Roughly the same amount 
of CPU time was used for all data points, the number of Monte Carlo steps per 
site thus scaling inversely with 
number of particles $N P$ of the pseudo-classical system. The whole run consisted of
numbers of subbatches ranging from 5 to 57 and the batch subaverages were used to 
estimate the approximate error bars of the total averages.

\section{Results and discussion}
\label{resdis}

Before discussing in detail the results for various quantities, we comment briefly on 
convergence of the PIMC scheme as a function of the Trotter number $P$. For different 
quantities, we have found considerably different convergence, the best case being that 
of quantities like, e.g., lattice 
constants, which are mainly related to softer interactions (non-bonded interactions and
torsional terms). In such cases, where the results obtained with different values of 
Trotter number $P$ were identical within the statistical error, the quantum limit was 
practically reached and no extrapolation was necessary. A considerably slower 
convergence is found for quantities like the energy, which depend crucially on 
fluctuations of degrees of freedom related to strong (bonded) 
interaction potentials. In these cases, an extrapolation to $P \rightarrow \infty$ was
performed in order to recover the true quantum values. We have used the standard formula 
\cite{tan}
\begin{equation}
A_{P} = A_{\infty} + {{a} \over {P^{2}}} + {{b} \over {P^{4}}} + O({{1} \over {P^{6}}}) \, ,
\end{equation}
which requires data for three different values of Trotter number $P$ in order to find the
extrapolated value $A_{\infty}$. 

In the discussion, we concentrate mainly on the comparison of quantities obtained from the 
quantum simulation to their classical counterparts, presented in \cite{rm}.
We start with the local quantities, in particular fluctuations of the 
internal coordinates, for which the quantum effects are found to be most pronounced. 

In Figs.~\ref{figdrcc} and \ref{figdrch}, we show the temperature dependence of the average
bond length fluctuation for C-C and C-H bonds, respectively. The distinguishing feature of 
quantum results for such fluctuations is their saturation at rather large values at low 
temperatures, instead of the classical vanishing 
($\sqrt{\langle (\delta r)^2 \rangle} \propto \sqrt{T}$ as $T \rightarrow 0$ in 
the classical case). In particular in case of the C-H bond, there is
a marked Trotter dependence of the results. By means of the above described extrapolation, we
found at $T = 50$ K the values of 0.05 $\AA$ for 
$\sqrt{\langle (\delta r_{CC})^2 \rangle}$ and 0.079 $\AA$ for 
$\sqrt{\langle (\delta r_{CH})^2 \rangle}$, which represent about 3 \% 
and 7 \% of the respective equilibrium bond length. These values are 
representative of the 
ground state values, since both curves appear to be entirely flat up to room temperature,
reflecting high frequencies of corresponding bond stretching phonon modes. In 
Figs.~\ref{figdccc} and \ref{figdhch}, average fluctuations of the bond angles 
$\theta_{CCC}$ and $\theta_{HCH}$ are shown as a function of temperature. Trotter 
extrapolation at $T = 50$ K yields here the values of 3.46$^{\circ}$ and 8.44$^{\circ}$ 
for $\sqrt{\langle (\delta \theta_{CCC})^2 \rangle}$ and
$\sqrt{\langle (\delta \theta_{HCH})^2 \rangle}$, respectively. While the former curve 
increases at room temperature by about 0.5$^{\circ}$ with respect to the ground state value,
the latter one corresponding to the bond angle involving two hydrogen atoms is completely 
flat.

In Fig.~\ref{figdcccc}, we show the temperature dependence of the average torsional angle 
fluctuation $\sqrt{\langle \phi_{CCCC}^2 \rangle}$ from the trans minimum, which already in 
the classical case \cite{rm} has been shown to play an important role in the physics of the 
system. The values at
$T = 50$ K and $T = 25$ K are already very close to each other, and therefore the $T = 25$ K
value of 5.59$^{\circ}$ can be considered to be representative of the ground state. The 
characteristic 
temperature, below which the difference between the classical and quantum values starts to 
increase rapidly, can be estimated to be about 150 K. At room temperature, the difference
is still about 0.5$^{\circ}$.

The internal energy per unit cell of the system is shown in Fig.~\ref{figenergy}. 
The dependence on $P$ is
in this case particularly pronounced, since the dominant contribution at low temperatures 
comes from the hard degrees of freedom which require larger values of the Trotter 
number in order to converge. The extrapolation to $P \rightarrow \infty$ is thus absolutely 
necessary here, and has been performed for all temperatures where data for three values of 
$P$ are available, i.e. 50, 100, 200 and 300 K. The extrapolated values at 50 K and 100 K 
suggest that the energy in this region would still somewhat decrease by approaching zero 
temperature and the extrapolated value of 62.125 kcal/mol at 50 K can be considered as 
an upper estimate 
of the ground state total energy. By subtracting the classical ground state energy which is
found to be equal to -7.39 kcal/mol per unit cell, we find a value of
17.38 kcal/mol per CH$_2$ group, which agrees well with the zero-point energy 
of 17.598 kcal/mol, 
obtained in Ref.~\cite{kdg} within quasi-harmonic approximation for a different force field.

Concerning the structural stability, for the smaller system we observed
at $T = 300$ K an occasional rotation of a whole chain from the "herringbone" 
structure, just as in the classical case \cite{rm}. No such rotation was observed
for the larger system, however. 

Before passing to the discussion of the temperature dependence of the lattice constants
$a$, $b$ and $c$, we would like to make a general comment on the statistical error of 
lattice constants of crystals evaluated within constant pressure PIMC scheme.
The statistical error $\sigma(\langle a \rangle_{run})$ of the lattice constant 
$\langle a \rangle_{run}$ averaged over a run consisting of $N$ configurations is given by 
\begin{equation}
\sigma(\langle a \rangle_{run}) = {{\sigma(a)}\over{\sqrt{{{N}\over{s}}}}} \; , 
\end{equation}
where $\sigma(a)$ is the intrinsic fluctuation of the quantity $a$, and $s$ is the 
corresponding statistical inefficiency, expressing the effect of correlations between
subsequent configurations of the Monte Carlo run \cite{allen}. In order to keep the Trotter 
error 
roughly constant at different temperatures, one usually keeps the product $P T$ constant.
For given system size and amount of CPU time, the number of configurations $N$ is inversely 
proportional to the Trotter number $P$ and thus directly proportional to temperature $T$.
On the other hand, the fluctuation $\sigma(a)$ of the lattice constant $a$ is essentially 
a fluctuation 
of the linear size of the system which is a purely classical fluctuation, and obeys the
equipartition theorem. Provided the 
elastic constants of the system do not vary too much, which should be well satisfied at low
temperatures, the relation $\sigma^{2}(a) \sim T$ should hold.
Combining the expressions together, we find 
\begin{equation}
\sigma(\langle a \rangle_{run}) \sim {{\sqrt{T}}\over{\sqrt{{{T}\over{s}}}}} \sim \sqrt{s} 
\; , 
\end{equation}
where the explicit dependence on $T$ in numerator and denominator has cancelled. 
Of course, there is still an implicit dependence hidden in the factor $s$, which increases 
with decreasing temperature because of growing system size in the Trotter direction. 
Nevertheless, in case of such purely classical fluctuation which vanishes as 
$T \rightarrow 0$, the situation is more favorable with respect to the case of 
a general fluctuation which would instead tend to a finite zero-point quantum value. 
In our results for the lattice constants, this is illustrated by the fact that the error bars
of points at lowest temperatures are not larger than those of points at higher temperatures.

The temperature dependence of the lattice constant $c$ is shown in Fig.~\ref{figc}. 
In both, classical and quantum cases, a lattice contraction with increasing temperature 
is observed. An interesting feature here is that at 
$T \sim 50$ K, the classical and quantum curves cross. At higher temperatures, the 
quantum result stays above the classical one while falling slightly below that for 
$T < 50$ K, where the quantum flattening appears. This behavior suggests the 
presence of at least two distinct quantum effects. Since in the classical case \cite{rm}
it turned out that there is a linear dependence between $c$ and 
$\langle \phi_{CCCC}^{2} \rangle$, we have tried the same plot for our quantum data,  
Fig.~\ref{figcvsphi2}, in order to separate different contributions.
In the latter plot, the classical and quantum curves are roughly parallel to each other, 
which shows that the low-temperature flattening of $c$ is a consequence of 
freezing of the thermal contribution to the torsional fluctuations. With the exception of 
the lowest temperatures, the shift of the quantum curve with respect to the classical one
appears to be temperature independent, being roughly equal to $0.0015 \AA$. This represents 
a zero-point expansion of the 
lattice along the $c$-direction arising from hard modes which are not significantly 
excited even at room temperature. 
While in the classical case all points are found to fall very well on a straight 
line in the whole temperature range \cite{rm}, in the quantum case a distinct 
upward bending of the curve is apparent at low temperatures.
To understand its origin, one might try to use the full formula 
\begin{equation}
c = c_0 \left( 1 - {{1}\over{4}} \sin^2{{\alpha}\over{2}} {{1}\over{2}}
\langle (\phi_{0} - \phi_{1})^{2} \rangle \right) 
= c_0 \left[ 1 - {{1}\over{4}} \sin^2{{\alpha}\over{2}} \langle \phi_{0}^{2} \rangle
\left(1 - {{\langle \phi_{0} \phi_{1}\rangle} \over {\langle \phi_{0}^{2}\rangle}} \right) 
\right] \; , \label{haeg}
\end{equation}
derived in 
Ref.~\cite{hagele} (see also Appendix B), which relates the contraction of the lattice 
constant $c$ to correlation functions of two neighboring torsional angle fluctuations 
$\phi_{0},\phi_{1}$. The temperature dependence of the normalized correlation function 
${{\langle \phi_{0} \phi_{1}\rangle} \over {\langle \phi_{0}^{2}\rangle}}$ is shown in 
Fig.~\ref{figctor1}. 
In contrast to the classical curve, which is flat in the whole temperature region, 
the quantum one is seen to increase in value at low temperatures where it becomes almost
a factor of two larger than the classical one. In Fig.~\ref{figcvsctor1}
we plot $c$ vs. ${{1}\over{2}} \langle (\phi_{0} - \phi_{1})^{2} \rangle$. In this plot, 
the low temperature upward bending from Fig.~\ref{figcvsphi2} has become less pronounced
and the points now exhibit a clear linear dependence, which suggests that the bending has 
its origin in the quantum reinforcement of the correlation function 
${{\langle \phi_{0} \phi_{1}\rangle} \over {\langle \phi_{0}^{2}\rangle}}$ of neighboring 
torsions at low temperatures. The slope of $1.275 \times 10^{-4} \AA$ deg$^{-2}$, 
however, turns out to be larger in magnitude by a factor of two with respect to the value of 
${{1}\over{4}} c_0 \sin^2{{\alpha}\over{2}} ({{\pi}\over{180}})^2 = 6.22 \times 10^{-5}\AA$ 
deg$^{-2}$ resulting from Eq.~(\ref{haeg}) (we took the values $c = 2.53 \AA$ and $\alpha =
180^{\circ} - 110.75^{\circ}$). About the same discrepancy
in the slope is found also for the classical data, and this has
been pointed out already in Ref.~\cite{rm} (where, however, the correlation function
between neighboring torsions was neglected and the discrepancy thus turned out to be 
just about a factor of 1.5). While in the classical case the thermal contraction could be 
modified
due to contribution of modes other than torsions, in the quantum case such harder modes are
mostly frozen even at room temperature, and do not contribute considerably. Thus in the 
quantum case the formula (\ref{haeg}) should yield a better agreement with the simulation. 
As it turns out, however, it captures qualitatively correctly the basic role of torsional
correlation functions in the thermal contraction, but fails in the quantitative aspect. 
We discuss the origin of this discrepancy in detail in the Appendix B. Analogously to the
classical case, no significant 
finite-size effects are seen on the lattice constant $c$. Comparing the quantum result 
to experimental data \cite{sl} in Fig.~\ref{figc}, we see that apart from the constant 
offset, the agreement has improved at low temperatures, due to the quantum flattening.

In Fig.~\ref{figa}, we show the temperature dependence of the lattice constant $a$.
At temperatures below 50 K, the quantum curve appears to be entirely flat, and the 
difference between the quantum and the classical result is in this region as large as
$0.13 \AA$, which represents a 2 \% effect. At all temperatures, the quantum curve lies
above the classical one. It is interesting to note that the difference between the 
two curves persists up to room temperature, being at $T = 300$ K equal to $0.06 \AA$, 
which is still about a half of the zero-temperature value. For this lattice constant, 
a very good agreement between the simulation and experimental results \cite{sl} is found, 
which proves that almost the whole low-temperature 
discrepancy between the classical simulation results and experiment is purely due to 
quantum effects. Similarly to the classical case, no finite-size effect is seen on the 
quantum curve at $T = 100$ K, while a small one is seen at $T = 300$ K. 

Analogously to the case of the lattice constant $c$, in Fig.~\ref{figavsdcccc} we plot 
the lattice constant $a$ vs. $\langle \phi_{CCCC}^{2} \rangle$. Interestingly,
in such a "scaling" plot, both classical and quantum results are found to collapse
nearly on the same straight line, which shows that the lattice constant $a$ does not
depend on temperature explicitly, but only implicitly, through the temperature dependence
of the torsional fluctuations. Since the latter dependence is very different in classical
and in quantum case, the behavior of $a$ is also substantially different.

In Fig.~\ref{figb}, the temperature dependence of the lattice constant $b$ is shown.
Similarly to the classical case, the error bars for $b$ are larger than those for $a$. 
The quantum flattening at low temperatures is now less pronounced, and in order
to find the limiting zero-temperature value of $b$, it would be necessary to go to even 
lower temperatures than 25 K. Here again, the quantum curve lies above the classical one
at all temperatures. At 25 K, the difference between them is about $0.03 \AA$, which 
represents a 0.6 \% effect, while at 300 K it decreases to about $0.02 \AA$. 
A finite-size effect very similar to the one in the classical case is observed here, too. 
It is visible already at 100 K, where the point for the system with C$_{24}$ chains falls
slightly below that for the system with C$_{12}$ chains, while being strongly pronounced at
300 K, where the difference is about $0.035 \AA$. The same finite-size effect is also
visible in the "scaling" plot in Fig.~\ref{figbvsdcccc}, where moreover at lower 
temperatures a downward bending of the quantum curve can be observed. Comparison to the 
experimental data \cite{sl} in Fig.~\ref{figb} in this case appears to be less good than in 
the case of $a$, however, for a detailed comparison simulation data for larger system sizes 
would be required.

We have also computed the diagonal elastic constants $c_{11},c_{22},c_{33}$ of the system.
Analogously to the classical case \cite{rm}, we evaluated $c_{11},c_{22}$ from the 
Parrinello-Rahman fluctuation formula \cite{pr}, 
\begin{equation}
c_{ik} = {{k_B T}\over{\langle V \rangle}} \langle e_i e_k \rangle^{-1}   \; ,
i,k = 1,2,3 \; ,
\label{prff}
\end{equation}
while for $c_{33}$ we used the Gusev-Zehnder-Suter formula \cite{gzs} 
in its approximate version suitable for small strain fluctuations
\begin{equation}
c_{ik} = -\sum_n \langle p_i e_n \rangle \langle e_n e_k \rangle^{-1}   \; ,
\label{newff}
\end{equation}
where $V$ is the super-cell volume and $p_i$ and $e_i$ are the diagonal components 
of the pressure tensor and strain tensor, respectively. This choice of methods was motivated
by the finding that in the classical case \cite{rm} significantly
smaller statistical errors resulted for $c_{33}$ from the Gusev-Zehnder-Suter formula 
(\ref{newff}), while for the other elastic constants errors were slightly smaller for the
Parrinello-Rahman fluctuation formula (\ref{prff}). Both formulae are classical and their 
use for evaluation of elastic constants also in case of PIMC technique is justified by the 
fact that strain fluctuations are classical objects which
have the same values in all Trotter slices. The results are shown on Figs.~\ref{figc11}, 
\ref{figc22} and \ref{figc33}. The data for $c_{11}$ exhibit 
a flattening at low temperatures and suggest that the ground state value is reduced with
respect to the classical one by about 2 GPa. A similar conclusion might be true also for 
$c_{22}$, where the large statistical error precludes a more detailed comparison. The best
results are obtained for $c_{33}$, analogously to the classical case. Here the flattening is
clearly seen and the ground state value is reduced due to quantum effects by about 20 GPa. In
Fig.~\ref{figc33gsvsdcccc2}, $c_{33}$ is plotted against $\langle \phi_{CCCC}^{2} \rangle$. 
The quantum points fall close to the line of collapse of the classical points,
which indicates that a dominant part of the quantum softening of $c_{33}$ has its origin 
in the finite value of zero-point torsional fluctuations. Not surprisingly,
error bars of the quantum data are much larger compared to the corresponding classical ones.
Our results show that for a strongly anisotropic crystal, in might be possible to obtain 
fairly good results for some components of the tensor of elastic constants, while other
components might be much more difficult to compute. In any case, calculation of elastic
constants within the PIMC scheme is at present computationally very demanding.

Finally, we have also studied isotope effects by simulating deuterated PE. In this case, we 
have 
performed the simulation only at the lowest temperature $T = 25$ K, with the same system size
and Trotter number as in case of normal (hydrogenated) polyethylene. The results for some
quantities are summarized in Table \ref{t1}. All three lattice constants are shorter in 
deuterated PE. The largest effect is 
seen on the lattice constant $a$, while in case of $b$ its relative magnitude is smaller by 
a factor of 3 and in that of $c$ by a factor of 20. 

\section{Conclusions}
\label{conc}

In this paper, we have demonstrated that for system sizes of several hundred atoms,
PIMC is a practical method 
allowing a fully quantum simulation of crystalline systems with many different energy
scales, like realistic explicit atom models for polymer crystals with no
constraints on the degrees of freedom. Even in the low-temperature region, where the 
system is close to its ground state, it is possible to calculate lattice constants 
and internal coordinates with a fairly good accuracy. On the other hand, an accurate
determination of elastic constants is still very difficult. The limitations of the method in
a study of the thermal expansion of the PE crystal are mainly set by the fact that the 
finite-size effects in the quantum case are more pronounced with respect to the classical
one, while at the same time it is more difficult to simulate larger 
systems, because of the extra Trotter dimension. It would be very helpful to have
for an anisotropic crystal a combined finite-size scaling scheme, allowing a 
simultaneous extrapolation of lattice constants to thermodynamic limit and Trotter
limit, analogous to the one developed in Ref.~\cite{mueser} for the specific heat of 
a cubic crystal. A prerequisite for such scaling, however, is an availability of 
high-accuracy data. A possible route here might be an improvement on the primitive 
PIMC algorithm by using a better approximation to the density matrix, similar to that
used for liquid $^4$He in Ref.~\cite{ceperley}, allowing a substantial reduction of 
the Trotter number. In a polymer crystal, the stiffest parts of the potential are 
the bond stretching terms, which formally have a form of pair interactions between
neighboring atoms. For such pair interactions, it is possible to calculate the exact
two-body density matrix, either by means of expansion in eigenfunctions, or by matrix
squaring. This exact form could be tabulated and used in the simulation, while the
rest of the potential would be treated in a standard way. Such a trick can be expected
to considerably improve the Trotter convergence, which in turn would enable simulation
of larger systems and increase the statistical accuracy of the results.

A detailed comparison to a quasi-harmonic approximation will be done in a forthcoming 
paper \cite{rmpb}. It would also be of interest to perform such an approximation for finite 
lattices in order to clarify the physical origin of the finite-size effects in both classical
and quantum cases, which still remain to be understood.

	Concerning the physics of the system, we have demonstrated that both in the
classical and the quantum case, the torsional fluctuations play a central role in the 
thermal expansion of the system. It would be desirable to have an analytical theory of
the lateral thermal and zero-point expansion, allowing to understand the origin of the
anisotropy in both cases. We have determined also some local quantities, in particular
zero-point fluctuations of internal coordinates, like bond lengths and bond and
torsional angles, knowledge of which might be of relevance for local experimental 
techniques. 

Finally, we would like to make a few remarks concerning modeling of polymer crystals
in general, taking into account quantum effects. It seems to us that for this sort 
of systems, force field 
building lags somewhat behind the development of simulation methods, concerning in 
particular the ability to reproduce anharmonic effects correctly. As already pointed 
out in Ref.~\cite{rm}, in the classical limit, where in principle all phonon modes can 
contribute to the thermal expansion of the system, there is a substantial difference 
between the properties of SLKB force field \cite{slkb} and KDG force field \cite{kdg}, 
as far as the thermal expansion coefficient $\alpha_3$ is concerned. In case of the former 
one, $\alpha_3$ is classically negative at all temperatures, and originates dominantly from
torsional fluctuations, while for the latter one it vanishes as $T \rightarrow 0$
\cite{rutledge}, which points to a large contribution of other modes. Such behavior 
might perhaps have to do with the equilibrium values in the bond stretching and angle 
bending terms since for 
the KDG force field these are substantially different from their average values, 
which in our opinion does not seem to be sufficiently justified. This also demonstrates
that the properties of 
a polymer crystal are much more sensitive to the details of the force field than those
of a liquid. In order to have a systematic control of the important anharmonic 
properties of a solid polymer system, an improvement on the side of the force field 
building is necessary. Such improvement would allow to explore fully the potential of 
existing classical and quantum simulation methods. One first and relatively simple 
thing to do in order to develop force fields suitable for quantum simulations would be
to use a quantum quasi-harmonic approximation to determine the ground state structure, 
instead of bare energy minimization, for fitting force fields to experimental 
structures. Obviously, a prerequisite for this 
is a better experimental characterization of the system in the whole range of
temperatures, making use of up-to-date experimental techniques, like, e.g., x-ray
diffraction with synchrotron sources. Such techniques should also allow a precise 
experimental determination of isotope effects on lattice constants and thermal
expansion, which might in principle help to decide which of the abovementioned force 
fields provides a better description of the real PE crystal. While for the SLKB force 
field the difference between classical and quantum value (the isotope effect is also
related to this difference) of $\alpha_3$ tends to vanish in the temperature region 
over 200 K, where the torsional fluctuations become thermally activated, in case of the 
KDG force field a large difference persists up to room temperature \cite{rutledge}. Accurate 
measurements of all components of the tensor of elastic constants in a wide region of 
temperatures would also be very interesting and helpful.

\acknowledgements

We would like to acknowledge stimulating discussions with P. C. H\"{a}gele, P. Nielaba 
and D. Ceperley, as well as correspondence with G. C. Rutledge and 
R. A. Stobbe. \\

\centerline{\bf APPENDIX A} 

In this appendix, we present the full form of the force field used in our present quantum
and previous classical simulation \cite{rm}, which is a slightly modified version of the 
SLKB force field \cite{slkb}. While the formal modifications have been
described in detail in Ref.~\cite{rm}, here we provide the actual numerical values of all 
parameters. The force field consists of bonded and non-bonded interactions.
The bonded interactions involve bond stretching, angle bending and torsions, as well as 
off-diagonal, or cross, terms, coupling together the different internal 
coordinates of the chains. The corresponding terms have the form of the 
following expressions:\\
a) Bond stretching, applying to all C-C and C-H bonds.
\begin{equation}
U(r) = {{1}\over{2}} k^{IJ} (r - r_0)^2
\end{equation}
b) Angle bending, applying to all C-C-C, C-C-H and H-C-H angles.
\begin{equation}
U(\theta) = {{1}\over{2}} k^{IJK} (\cos\theta - \cos\theta_0)^2
\end{equation}
c) Torsion terms, applying to all C-C-C-C, C-C-C-H and H-C-C-H sequences, 
\begin{equation}
U(\varphi) = {{1}\over{2}} V^{IJKL} (1 + \cos3\varphi) \; ,
\end{equation}
where $\varphi$ is the torsional angle ($\varphi = 0$ corresponds to cis and
$\varphi = \pi$ corresponds to trans). Throughout the rest of the paper, we use
also torsional angle $\phi$ defined as fluctuation of $\varphi$ from the trans
value, $\phi = \varphi - \pi$.\\
d) Bond-angle and bond-bond cross terms, applying to all C-C-C, C-C-H and H-C-H
angles, 
\begin{eqnarray}
U(r_1,r_2,\theta) &=& 
k_{r_1,\theta}^{IJK} (r_1 - r_{10}) (\cos\theta - \cos\theta_0) +
k_{r_2,\theta}^{IJK} (r_2 - r_{20}) (\cos\theta - \cos\theta_0) \nonumber \\
&+& k_{r_1,r_2}^{IJK} (r_1 - r_{10}) (r_2 - r_{20}) \; ,
\end{eqnarray}
where $r_1$ and $r_2$ are the bond lengths of the I-J and J-K bonds adjacent
to an IJK bond angle $\theta$.\\
e) One center angle-angle cross terms, applying to all pairs of bond angles
about a tetrahedral carbon atom sharing a common bond, 
\begin{equation}
U(\theta_1,\theta_2) = G^{IJ:KL} 
(\cos\theta_1 - \cos\theta_t) (\cos\theta_2 - \cos\theta_t) \; ,
\end{equation}
where $\theta_1,\theta_2$ are the JIK and JIL bond angles, respectively,
and I is the tetrahedral carbon. Here, $\theta_t$ is the tetrahedral angle.\\
f) Two-center angle-angle cross terms, applying to all C-C-C-C, C-C-C-H and 
H-C-C-H sequences, 
\begin{equation}
U(\varphi,\theta_1,\theta_2) = F^{I:JK:L} \cos \varphi
(\cos\theta_1 - \cos\theta_t) (\cos\theta_2 - \cos\theta_t) \; ,
\end{equation}
where $\varphi$ is the torsional angle corresponding to the sequence IJKL,
$\theta_1,\theta_2$ are the IJK and JKL bond angles and $\theta_t$ is again
the tetrahedral angle. 
Numerical values of all the parameters in the above 
expressions are contained in the table Tab.\ref{t2}. 

The non-bonded interaction has the form $U(r) = A e^{-B r} - C r^{-6}$ between 
atoms on different chains and atoms on the same chain separated by more than 
two bonds (1 -- 2 and 1 -- 3 interactions are excluded).
Numerical values of all the parameters for all three pairs of atoms (C-C, H-H
and C-H) are contained in the table Tab.\ref{t3}. Details concerning cutoff and long-range
corrections can be found in Ref.~\cite{rm}. \\

\centerline{\bf APPENDIX B} 

In this appendix we discuss in detail the origin of the quantitative discrepancy between
the slope of the $c$ vs. ${{1}\over{2}} \langle (\phi_{0} - \phi_{1})^{2} \rangle$ line
as obtained from formula (\ref{haeg}) and from our quantum simulation data 
(Fig.~\ref{figcvsctor1}). To this end, we have to look more closely to the way the formula 
has been derived in Ref.~\cite{hagele}. The model used assumes a single chain consisting of 
rigid C-C bonds with length $a$ and rigid C-C-C angles $\theta_{CCC} = \pi - \alpha$, subject
to torsional deformations only. The expression (\ref{haeg}) then evaluates $c$ simply as 
the average end-to-end distance between carbon atoms separated by four bonds.
Let us denote by $\vec{r},\vec{r}^{'}$ the instantaneous positions of the two atoms, and by
$\vec{r}_{0},\vec{r}_{0}^{'}$ their equilibrium positions around which they oscillate with
instantaneous fluctuations $\vec{\Delta},\vec{\Delta}^{'}$. The squared end-to-end 
distance is then given by 
\begin{equation}
(\vec{r} - \vec{r}^{'})^2 = 
(\vec{r}_{0} - \vec{r}_{0}^{'} + \vec{\Delta} - \vec{\Delta}^{'})^2 = 
(\vec{r}_{0} - \vec{r}_{0}^{'})^2 + (\vec{\Delta} - \vec{\Delta}^{'})^2 \; ,
\end{equation}
where the cross term has vanished, because the fluctuations $\vec{\Delta},\vec{\Delta}^{'}$ 
are orthogonal to $\vec{r}_{0} - \vec{r}_{0}^{'}$ (since the bond lengths and angles are 
assumed to be rigid, the only possible small displacements of carbon atoms are those 
perpendicular to the plane of the unperturbed all-trans chain). This expression contains 
apart from the squared distance between the equilibrium positions 
$(\vec{r}_{0} - \vec{r}_{0}^{'})^2$, which is directly
related to the lattice constant $c$, also a fluctuation term. Neglecting this term for 
a relatively short segment of chain, as done in Ref.~\cite{hagele}, results in 
underestimating of the thermal contraction. 
In order to improve on this, one has to consider a longer segment of the chain, which would,
however, require a knowledge of 
correlation functions between torsions displaced by more than one bond. This is clearly 
impossible within the simple single-chain model assumed in Ref.~\cite{hagele}, where the 
crystal environment of the chain is neglected entirely. This together with a
separable torsional potential classically leads to vanishing of all correlation 
functions between displaced torsions, and in turn to coiling of longer segments 
of a chain. In a crystal it is just the external non-bonded field of all other chains acting
directly on the absolute coordinates of the atoms of a given chain 
(rather than on the torsional angles, which by their very nature have a character of 
relative coordinates), which induces 
non-zero correlations between the torsions in the chain. These correlations reflect the
existence of the underlying 3D crystal structure with its translational long-range
order and result in an overall coherent shortening of the chain, instead of its coiling.
The quantitative agreement of the formula with experiment, as stated in Ref.~\cite{hagele}, 
thus appears to be accidental and arises due to compensation of two effects: neglecting the 
crystal field contribution lowers the effective torsional constant by a factor of two which 
in turn increases the torsional fluctuations and compensates for the lower value of the
proportionality constant in the expression (\ref{haeg}). 

Now we derive a few formulas similar to (\ref{haeg}), taking into account progressively longer
segments of a chain. In general, we are interested in calculating the end-to-end distance of
a segment of a chain consisting of $n=2m$ bonds. The corresponding vector can be expressed 
as \cite{flory}
\begin{equation}
\vec{R_n} = \sum_{k=1}^{n} \prod_{i=1}^{k} T_{i}  \;\; (a,0,0)^{T} \;,
\end{equation}
where the $3 \times 3$ matrices $T_{i}$ are defined as follows
\begin{equation}
T_{i} = \left|
\begin{array}{ccc}
\cos{\alpha} & \sin{\alpha} & 0  \\
\sin{\alpha} \cos{\phi_i} & -\cos{\alpha} \cos{\phi_i} & \sin{\phi_i} \\
\sin{\alpha} \sin{\phi_i} & -\cos{\alpha} \sin{\phi_i} & -\cos{\phi_i}
\end{array} \right| \; , \label{mat}
\end{equation}
angles $\phi_i$ being the torsional angles. Evaluating $\langle |\vec{R}_n| \rangle$ and 
keeping just terms up to second order in the angles $\phi_i$ we find the following 
approximations $c^{(m)} = {{1}\over{m}} \langle |\vec{R}_{2 m}| \rangle$ to the contracted 
lattice constant $c$ (the actual calculation has been performed by Mathematica)
\begin{eqnarray}
c^{(2)} &=& {{1}\over{2}} \langle |\vec{R}_{4}| \rangle = 
c_0 \mbox{\LARGE{[}} 1 - {{1}\over{4}} \sin^2{{\alpha}\over{2}} 
(\langle {\phi_0}^2 \rangle - \langle \phi_0 \phi_1 \rangle) \mbox{\LARGE{]}}  \nonumber \\
c^{(3)} &=& {{1}\over{3}} \langle |\vec{R}_{6}| \rangle = c_0 \mbox{\LARGE{[}}
1 - {{1}\over{9}} \sin^2{{\alpha}\over{2}} 
(4 \langle {\phi_0}^2 \rangle - 5 \langle \phi_0 \phi_1 \rangle
+ 2 \langle \phi_0 \phi_2 \rangle - \langle \phi_0 \phi_3 \rangle
) \mbox{\LARGE{]}} \nonumber \\
c^{(4)} &=& {{1}\over{4}} \langle |\vec{R}_{8}| \rangle = c_0 \mbox{\LARGE{[}}
1 - {{1}\over{16}} \sin^2{{\alpha}\over{2}} 
(10 \langle {\phi_0}^2 \rangle - 14 \langle \phi_0 \phi_1 \rangle
+ 8 \langle \phi_0 \phi_2 \rangle - 5 \langle \phi_0 \phi_3 \rangle
+ 2 \langle \phi_0 \phi_4 \rangle \nonumber \\
&-& \langle \phi_0 \phi_5 \rangle) \mbox{\LARGE{]}} \nonumber \\
c^{(5)} &=& {{1}\over{5}} \langle |\vec{R}_{10}| \rangle = c_0 \mbox{\LARGE{[}}
1 - {{1}\over{25}} \sin^2{{\alpha}\over{2}} 
(20 \langle {\phi_0}^2 \rangle - 30 \langle \phi_0 \phi_1 \rangle
+ 20\langle \phi_0 \phi_2 \rangle - 14\langle \phi_0 \phi_3 \rangle
+ 8 \langle \phi_0 \phi_4 \rangle \nonumber \\
&-& 5 \langle \phi_0 \phi_5 \rangle
+ 2 \langle \phi_0 \phi_6 \rangle - \langle \phi_0 \phi_7 \rangle
) \mbox{\LARGE{]}} \nonumber \\
c^{(6)} &=& {{1}\over{6}} \langle |\vec{R}_{12}| \rangle = c_0 \mbox{\LARGE{[}}
1 - {{1}\over{36}} \sin^2{{\alpha}\over{2}} 
(35 \langle {\phi_0}^2 \rangle - 55 \langle \phi_0 \phi_1 \rangle
+ 40\langle \phi_0 \phi_2 \rangle - 30 \langle \phi_0 \phi_3 \rangle
+ 20\langle \phi_0 \phi_4 \rangle \nonumber \\ 
&-& 14 \langle \phi_0 \phi_5 \rangle
+ 8 \langle \phi_0 \phi_6 \rangle - 5 \langle \phi_0 \phi_7 \rangle
+ 2 \langle \phi_0 \phi_8 \rangle - \langle \phi_0 \phi_9 \rangle
) \mbox{\LARGE{]}} \; , \label{corrtor}
\end{eqnarray}
where $c_0 = 2 a \cos{{\alpha}\over{2}}$ is the lattice constant of the unperturbed
chain.

We introduce now off-plane displacements $z_i$ of carbon atoms along local 
$z$-axes defined for each C atom by the unit vector 
$\vec{k}_i = (\vec{a}_i \times \vec{a}_{i+1})/a^2 \sin{\alpha}$,
where $\vec{a}_i$ is the vector connecting carbon atoms $i - 1$ and $i$.
Vectors $\vec{k}_i$ are perpendicular to the plane of the all-trans chain
and their directions alternate below and above the plane. It can then be easily 
shown \cite{stobbe} that the torsional angles $\phi_i$ defined via $\cos{\phi_i} 
= - (\vec{a}_{i-1} \times \vec{a}_{i}) \cdot (\vec{a}_i \times \vec{a}_{i+1})
/a^4 \sin^2{\alpha}$, or, alternatively, $\sin{\phi_i} = \vec{a}_{i-1} \cdot
(\vec{a}_{i} \times \vec{a}_{i+1})/a^3 \sin^2{\alpha}$, can be in first order
in displacements $z_i$ expressed as 
$\phi_i = (-z_{i-2} - z_{i-1} + z_{i} + z_{i+1})/a \sin{\alpha}$. Upon substitution
of this latter expression to (\ref{corrtor}), we find 
\begin{equation}
c^{(m)} = c^{(\infty)} + {{1}\over{2 m^2}} {{\langle (z_0 - z_{2m})^2 \rangle}\over{c_0}} \; ,
\end{equation} 
where 
\begin{equation}
c^{(\infty)} = c_0 \left( 1 - {{\langle (z_0 - z_2)^2 \rangle}\over{2 c_0^2}} \right) 
\end{equation} 
is the limiting $m \rightarrow \infty$ value of the contracted lattice constant $c$. 
The last result has been derived also in Ref.~\cite{stobbe} by means of a simple 
geometrical argument, expressing $c$ as a projection of the segment consisting of
two C-C bonds, whose end-to-end length is $c_0$, on the plane of the unperturbed chain.

Applying the formula for $c^{(3)}$ corresponding to 
a segment consisting of six C-C bonds to our simulation data, we find a slope which is 
still larger by about 50 \% than the theoretical one. A characteristic property of all the 
expressions (\ref{corrtor})
is that the coefficients of torsion correlation functions increase in magnitude with 
the length of the segment. The correct limiting result for thermal contraction 
thus emerges progressively as a consequence of a delicate cancellation between the terms, 
which means that a knowledge of correlation functions at many different displacements 
with very good accuracy would be required. Such behavior reflects the lack of existence of 
a preferred crystal direction for a chain in the formulation employing exclusively torsional
angles. In contrast to this, upon substitution of the
true off-plane displacements of the atoms into these expressions, the limiting exact 
result emerges in a very simple form, which has a clear geometrical interpretation and 
contains just a single correlation function between second neighbors. 
This demonstrates that while the relation between the torsional fluctuations and the lattice
constant $c$ provides a very useful insight for qualitative understanding of the thermal 
contraction of $c$, it would be at the same time hard to push these considerations 
to a truly quantitative level.

\begin{figure}
\caption[]{Temperature dependence of the average fluctuation
$\sqrt{\langle (\delta r_{CC})^2 \rangle}$ of the C-C bond length, in classical
and quantum case, shown for different system sizes. In this and most of the 
following figures, statistical error bars are shown, lines are for visual help 
only. In all figures, the same symbol is used for quantum results corresponding
to different values of the Trotter number $P$. When at a given temperature the 
results for different values of $P$ are indistinguishable within the statistical
error, like in this figure, the Trotter numbers are not indicated explicitly.}
\label{figdrcc}
\end{figure}

\begin{figure}
\caption[]{Temperature dependence of the average fluctuation
$\sqrt{\langle (\delta r_{CH})^2 \rangle}$ of the C-H bond length, in classical
and quantum case, shown for different system sizes. When the quantum results at
a given temperature exhibit a pronounced dependence on the Trotter number $P$, 
like in this figure, the corresponding values of $P$ are indicated next to the 
symbols.}
\label{figdrch}
\end{figure}

\begin{figure}
\caption[]{Temperature dependence of the average fluctuation
$\sqrt{\langle (\delta \theta_{CCC})^2 \rangle}$ of the C-C-C bond angle, in classical and 
quantum case, shown for different system sizes.}
\label{figdccc}
\end{figure}

\begin{figure}
\caption[]{Temperature dependence of the average fluctuation
$\sqrt{\langle (\delta \theta_{HCH})^2 \rangle}$ of the H-C-H bond angle, in classical and 
quantum case, shown for different system sizes.}
\label{figdhch}
\end{figure}

\begin{figure}
\caption[]{Temperature dependence of the average torsional angle fluctuation
$\sqrt{\langle \phi_{CCCC}^2 \rangle}$ from the trans minimum, in classical and quantum 
case, shown for different system sizes.}
\label{figdcccc}
\end{figure}

\begin{figure}
\caption[]{Temperature dependence of the internal energy per unit cell of 
quantum system 
with C$_{12}$ chains, shown for different values of the Trotter number $P$, 
together with an extrapolation to $P \rightarrow \infty$. Note the strong 
dependence on the Trotter number $P$.} 
\label{figenergy}
\end{figure}

\begin{figure}
\caption[]{Temperature dependence of the lattice parameter $c$, in classical and quantum 
case, shown for different system sizes, together with the experimental data \cite{sl}.}
\label{figc}
\end{figure}

\begin{figure}
\caption[]{Lattice parameter $c$ vs. $\langle \phi_{CCCC}^2 \rangle$, in 
classical and quantum case, shown for different system sizes. Note that also in
the quantum case, the dependence is almost linear over the whole temperature 
range. The values of temperature, which is a parameter of the plot, are
indicated next to the symbols.}
\label{figcvsphi2}
\end{figure}

\begin{figure}
\caption[]{Temperature dependence of the normalized correlation function 
${{\langle \phi_{0} \phi_{1}\rangle} \over {\langle \phi_{0}^{2}\rangle}}$ of 
fluctuations of two neighboring torsional angles from the trans minima, in the
classical and quantum case, for the system with C$_{12}$ chains. Note the 
pronounced dependence on the Trotter number $P$.}
\label{figctor1}
\end{figure}

\begin{figure}
\caption[]{Lattice parameter $c$ vs. 
${{1}\over{2}} \langle (\phi_{0} - \phi_{1})^{2} \rangle$ in the quantum case 
for system with C$_{12}$ chains. The values of temperature, which is 
a parameter of the plot, are indicated next to the symbols.
The points can be fitted by a line $c = 2.5327 - 1.275 \times 10^{-4} 
{{1}\over{2}} \langle (\phi_{0} - \phi_{1})^{2} \rangle$.}
\label{figcvsctor1}
\end{figure}

\begin{figure}
\caption[]{Temperature dependence of the lattice parameter $a$, in classical and quantum 
case, shown for different system sizes, together with the experimental data \cite{sl}.}
\label{figa}
\end{figure}

\begin{figure}
\caption[]{Lattice parameter $a$ vs. $\langle \phi_{CCCC}^2 \rangle$, in 
classical and quantum case, shown for different system sizes. The values of 
temperature, which is a parameter of the plot, are indicated next to the 
symbols. Note the collapse of both classical and quantum results on almost the 
same straight line over the whole temperature range.}
\label{figavsdcccc}
\end{figure}

\begin{figure}
\caption[]{Temperature dependence of the lattice parameter $b$, in classical 
and quantum case, shown for different system sizes, together with the 
experimental data \cite{sl}. Note the strong finite-size effect at higher 
temperatures for both classical and quantum results.}
\label{figb}
\end{figure}

\begin{figure}
\caption[]{Lattice parameter $b$ vs. $\langle \phi_{CCCC}^2 \rangle$, in 
classical and quantum case, shown for different system sizes. The values of
temperature, which is a parameter of the plot, are indicated next to the
symbols. Note the strong finite-size effect at higher temperatures for both 
sets of data as well as the downward bending of the quantum curve at low 
temperatures.}
\label{figbvsdcccc}
\end{figure}

\begin{figure}
\caption[]{Elastic constant $c_{11}$ as a function of temperature, in both classical and
quantum case, shown for different system sizes.}
\label{figc11}
\end{figure}

\begin{figure}
\caption[]{Elastic constant $c_{22}$ as a function of temperature, in both classical and
quantum case, shown for different system sizes.}
\label{figc22}
\end{figure}

\begin{figure}
\caption[]{Elastic constant $c_{33}$ as a function of temperature, in both classical and
quantum case, shown for different system sizes.}
\label{figc33}
\end{figure}

\begin{figure}
\caption[]{Elastic constant $c_{33}$ vs. $\langle \phi_{CCCC}^2 \rangle$, 
in both classical and quantum case, for system with C$_{12}$ chains. The values
of temperature, which is a parameter of the plot, are indicated next to the 
symbols. Note that the quantum results fall close to the line passing through 
the classical results.}
\label{figc33gsvsdcccc2}
\end{figure}

\begin{table}
\begin{tabular}{|c|c|c|c|} \hline
  & normal PE               & deuterated PE        & difference \\ \hline
$a$ & $7.201  \pm 0.003  \AA$   & $7.170  \pm 0.003  \AA$ & -0.43 \%   \\ \hline
$b$ & $4.928  \pm 0.002  \AA$   & $4.921  \pm 0.002  \AA$ & -0.14 \%   \\ \hline
$c$ & $2.5297 \pm 0.0001 \AA$   & $2.5292 \pm 0.0001 \AA$ & -0.02 \%   \\ \hline
$\sqrt{\langle \phi_{CCCC}^2 \rangle}$ & $5.59 \pm 0.01 ^{\circ}$ & $5.26 \pm 0.015 ^{\circ} $ & -5.9 \%   \\ \hline
$\sqrt{\langle (\delta \theta_{HCH})^2 \rangle}$ & $8.28 \pm 0.001 ^{\circ}$ & $7.17 \pm 0.003 ^{\circ} $ & -13.4 \%   \\ \hline
$\sqrt{\langle (\delta r_{CH})^2 \rangle}$ & $0.074 \pm 0.00001 \AA$ & $0.065 \pm 0.00001 \AA$ & -12.1 \%   \\ \hline
\end{tabular}
\caption[...]{Isotope effect: deuterated PE compared to normal (hydrogenated) PE. All values
correspond to the system with C$_{12}$ chains and $P = 144$ at $T = 25$ K. No extrapolation 
to $P \rightarrow \infty$ has been performed here. }
\label{t1}
\end{table}

\begin{table}
\begin{tabular}{||c|c||c|c||} \hline
$r_{0}^{CC} [\AA]$ & 1.53 & $k^{HCH}_{r,\theta}$ [kcal/(mol$\AA$)]& 0 \\ \hline
$k^{CC}$ [kcal/(mol$\AA^2$)] & 617.058 & $k^{HCH}_{r1,r2}$ [kcal/(mol$\AA^2$)] & 0 \\ \hline
$r_{0}^{CH} [\AA]$ & 1.09 & $k^{CCH}_{r1,\theta}$ [kcal/(mol$\AA$)] & -34.0818 \\ \hline
$k^{CH}$ [kcal/(mol$\AA^2$)] & 654.455 & $k^{CCH}_{r2,\theta}$ [kcal/(mol$\AA$)] & 0 \\ \hline
$k^{HCH}$ [kcal/mol] & 76.952 & $k^{CCH}_{r1,r2}$ [kcal/(mol $\AA^2$)]& 0 \\ \hline
$\theta_{0}^{HCH} [^\circ]$ & 107.899 & $k^{CCC}_{r,\theta}$ [kcal/(mol$\AA$)] & -54.494 \\ \hline
$k^{CCH}$ [kcal/mol] & 85.726 & $k^{CCC}_{r1,r2}$ [kcal/(mol$\AA^2$)] & 25.9337 \\ \hline
$\theta_{0}^{CCH} [^\circ]$ & 109.469 & $G^{CC:CH}$ [kcal/mol]& -6.0034 \\ \hline
$k^{CCC}$ [kcal/mol] & 107.446 & $G^{CC:HH}$ [kcal/mol]& -3.6409 \\ \hline
$\theta_{0}^{CCC} [^\circ]$ & 110.999 & $G^{CH:CC}$ [kcal/mol]& 2.9127 \\ \hline
$V^{HCCH}$ [kcal/mol] & 0.2776 & $G^{CH:CH}$ [kcal/mol]& 0 \\ \hline
$V^{CCCH}$ [kcal/mol] & 0.2776 & $F^{H:CC:H}$ [kcal/mol]& -16.0398 \\ \hline
$V^{CCCC}$ [kcal/mol] & 0.2776 & $F^{C:CC:H}$ [kcal/mol]& -15.5343 \\ \hline
 & & $F^{C:CC:C}$ [kcal/mol]& -33.3664 \\ \hline 
\end{tabular}
\caption[...]{Parameters of the bonded interaction of the force field. The left column 
contains the coefficients of terms diagonal in the internal coordinates of the chains, 
while the right one contains the coefficients of the off-diagonal terms.}
\label{t2}
\end{table}

\begin{table}
\begin{tabular}{||c|c|c|c||} \hline
atoms & A [kcal/mol]   & B [$\AA^{-1}$] &      C [$\AA^6$ kcal/mol] \\ \hline 
C-C   &       14889.0  &      3.089     &     639.58 \\ \hline 
H-H   &        2640.2  &      3.739     &      27.39 \\ \hline 
C-H   &        4300.9  &      3.416     &     137.44 \\ \hline 
\end{tabular}
\caption[...]{Parameters of the non-bonded interaction of the force field.}
\label{t3}
\end{table}

\unitlength1mm

\begin{picture}(96,80)
\put(0,0){\psfig{figure=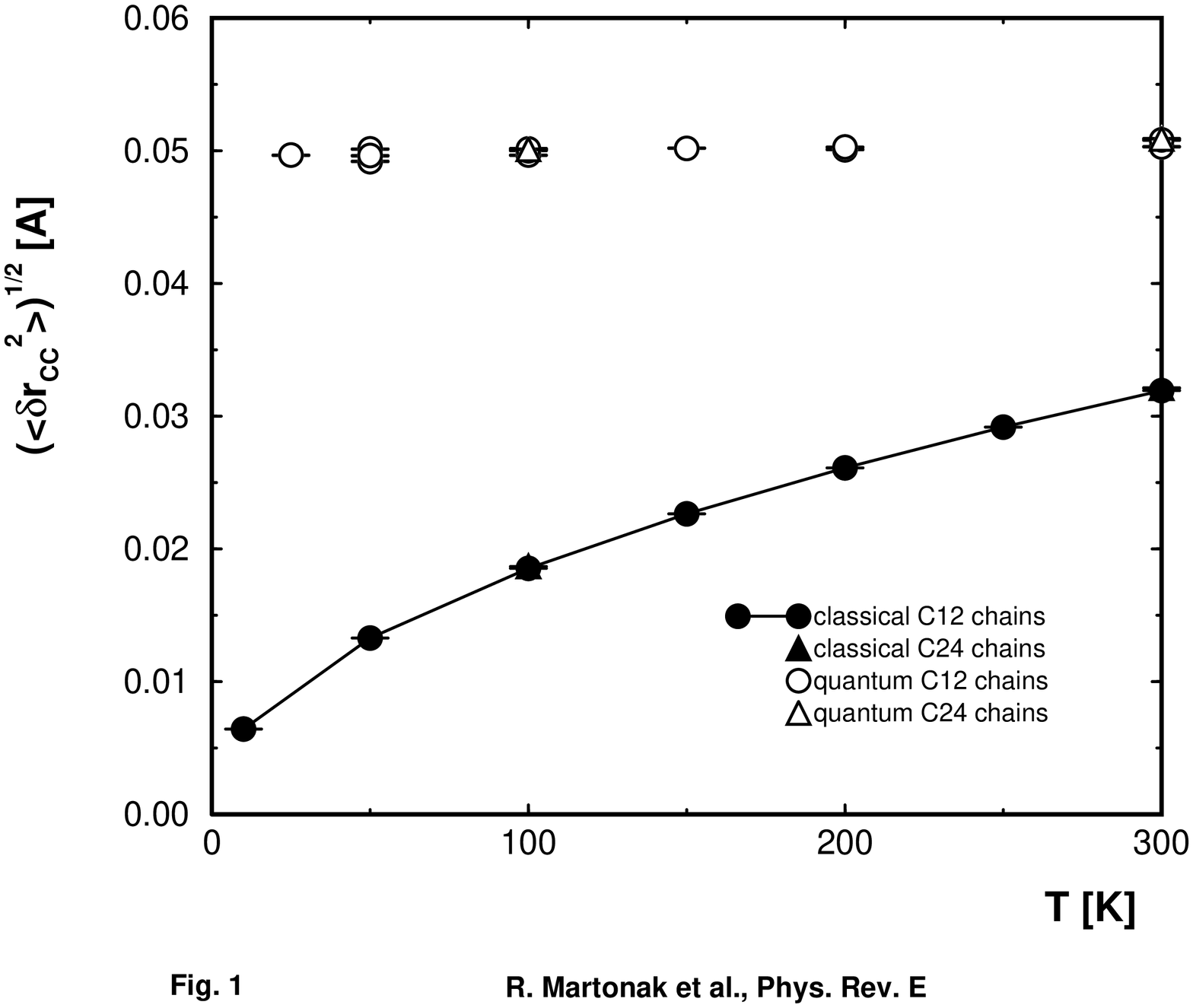,width=96mm,height=80mm,angle=-90}}
\end{picture}
\vskip 4.cm

\begin{picture}(96,80)
\put(0,0){\psfig{figure=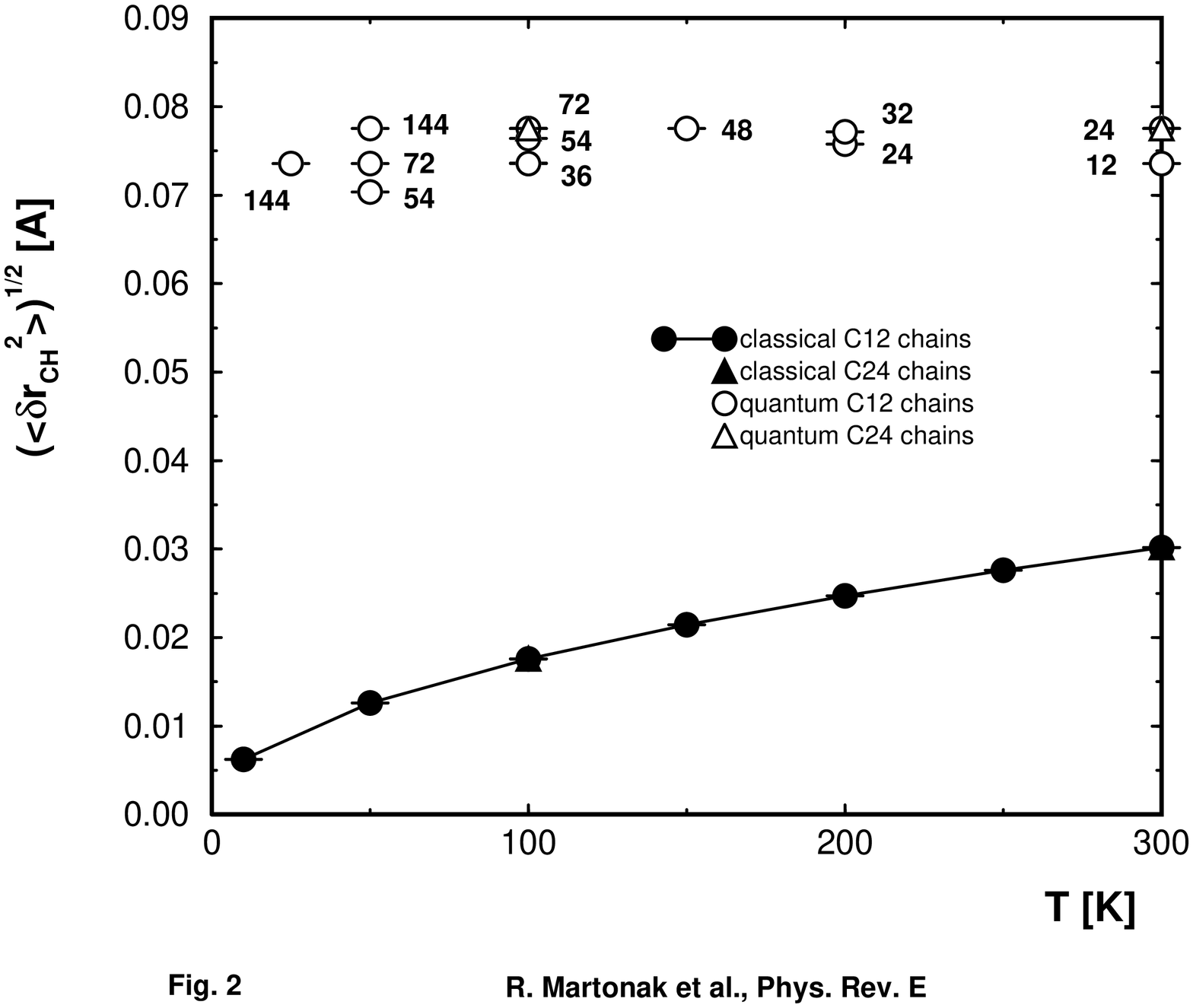,width=96mm,height=80mm,angle=-90}}
\end{picture}
\vfill\eject

\begin{picture}(96,80)
\put(0,0){\psfig{figure=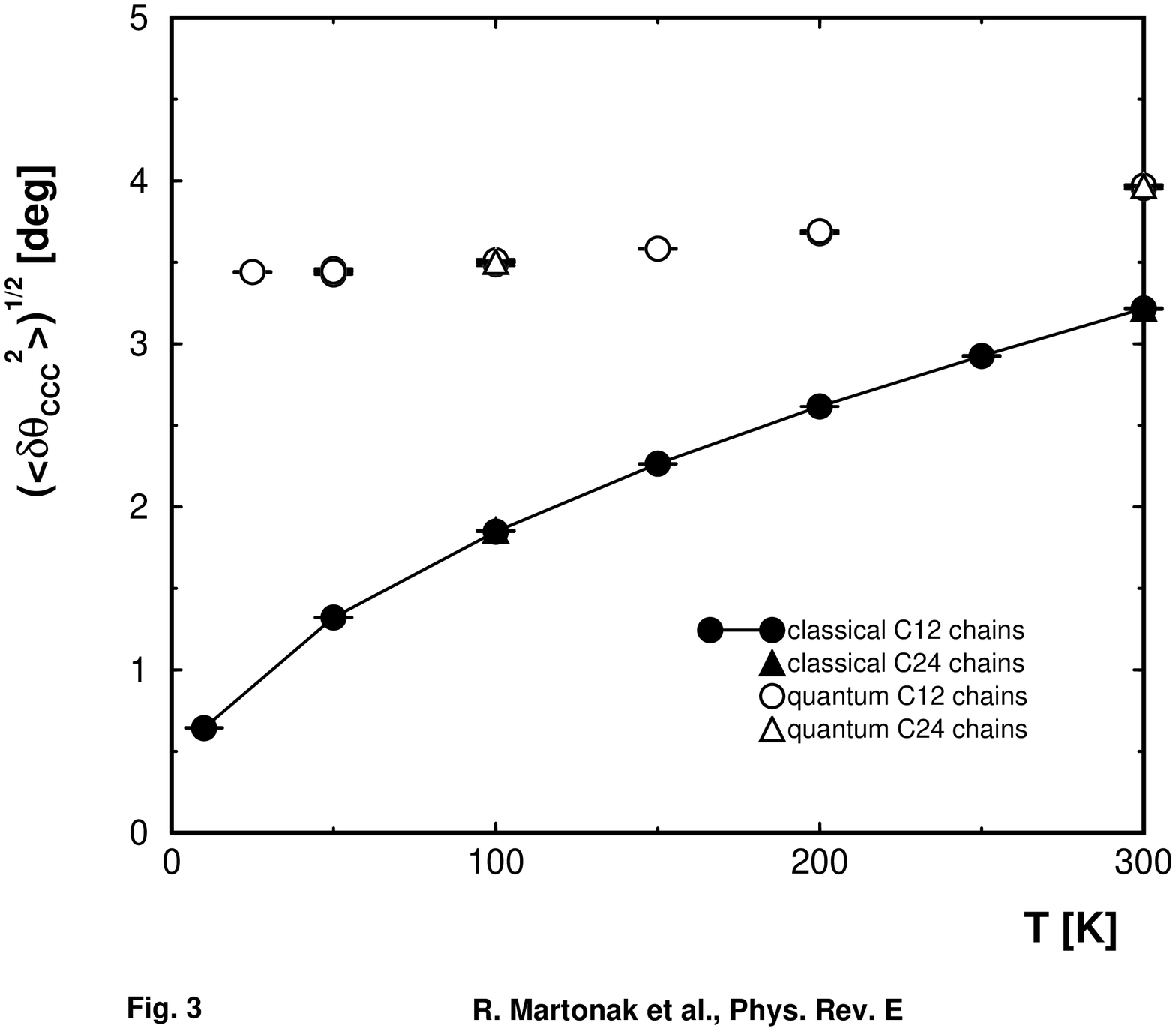,width=96mm,height=80mm,angle=-90}}
\end{picture}
\vskip 4.cm

\begin{picture}(96,80)
\put(0,0){\psfig{figure=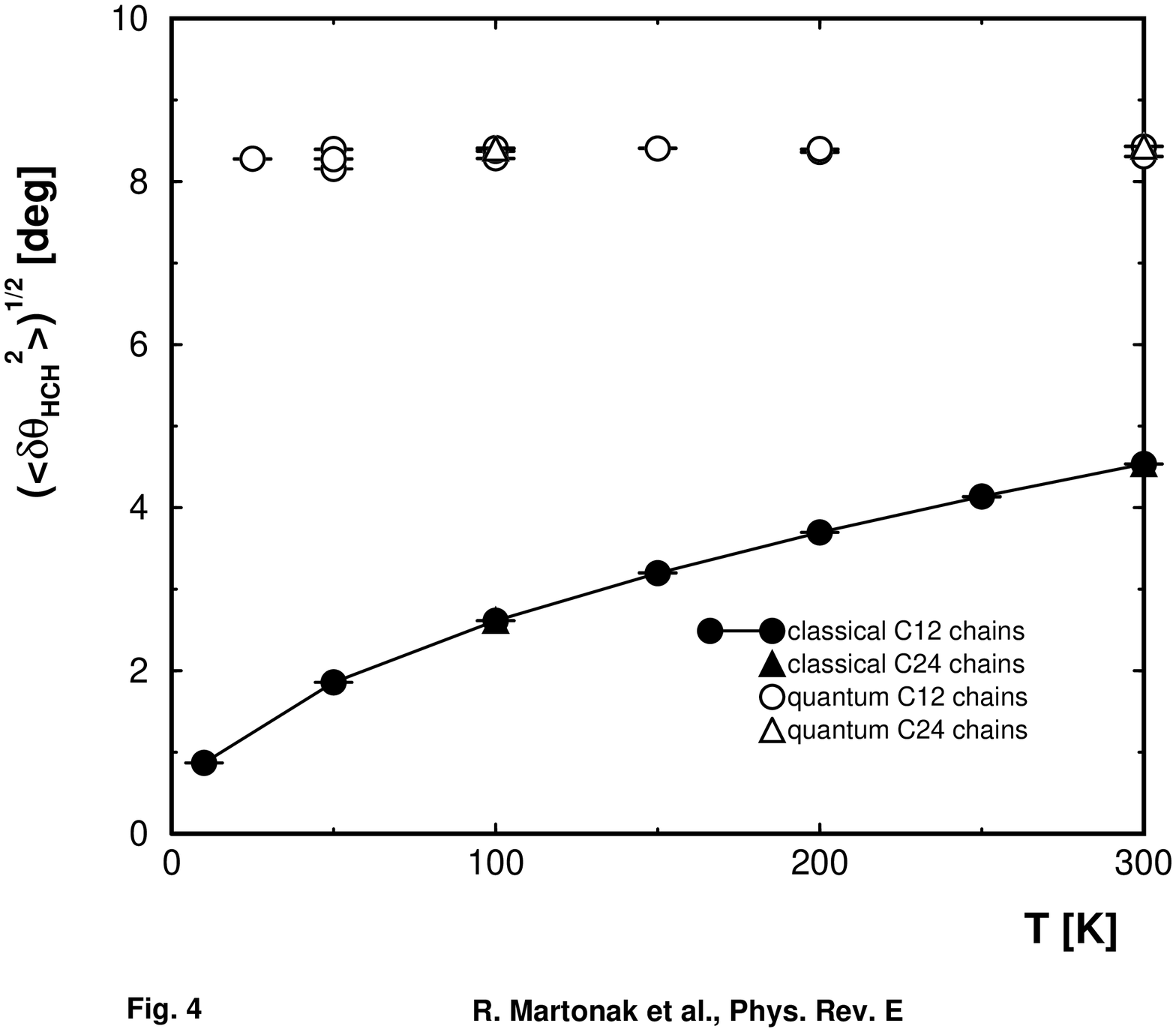,width=96mm,height=80mm,angle=-90}}
\end{picture}
\vfill\eject

\begin{picture}(96,80)
\put(0,0){\psfig{figure=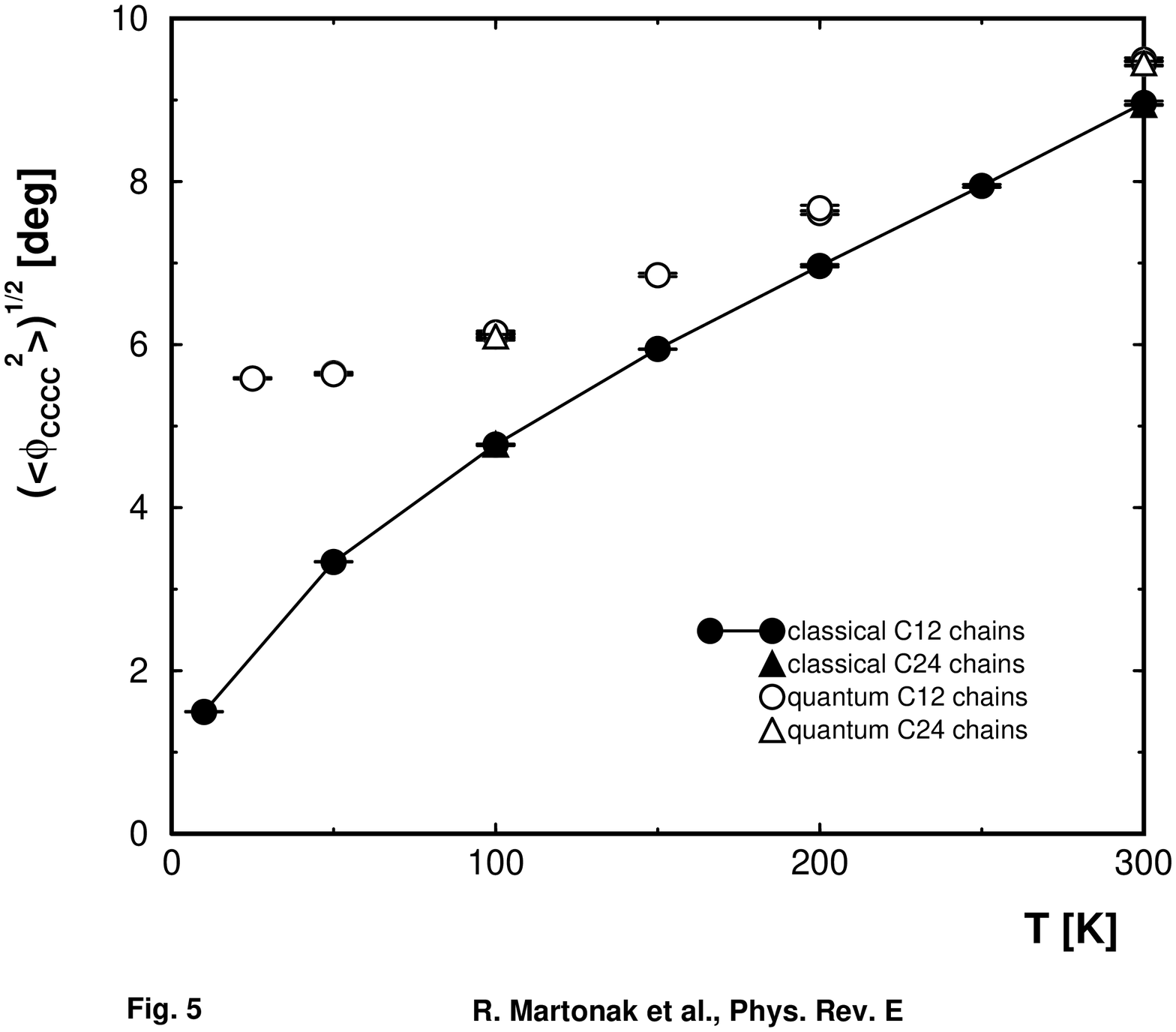,width=96mm,height=80mm,angle=-90}}
\end{picture}
\vskip 4.cm

\begin{picture}(96,80)
\put(0,0){\psfig{figure=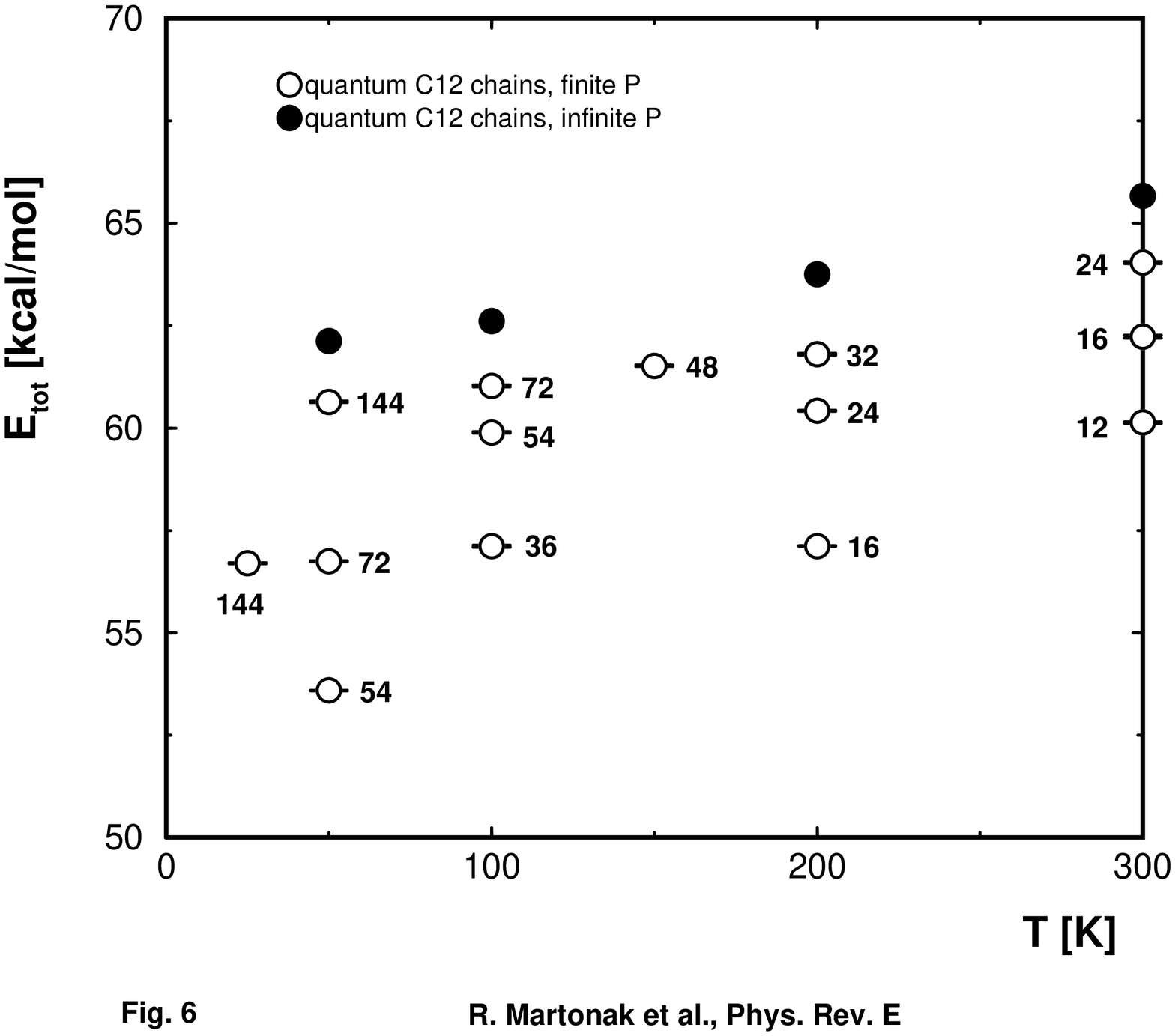,width=96mm,height=80mm,angle=-90}}
\end{picture}
\vfill\eject

\begin{picture}(96,80)
\put(0,0){\psfig{figure=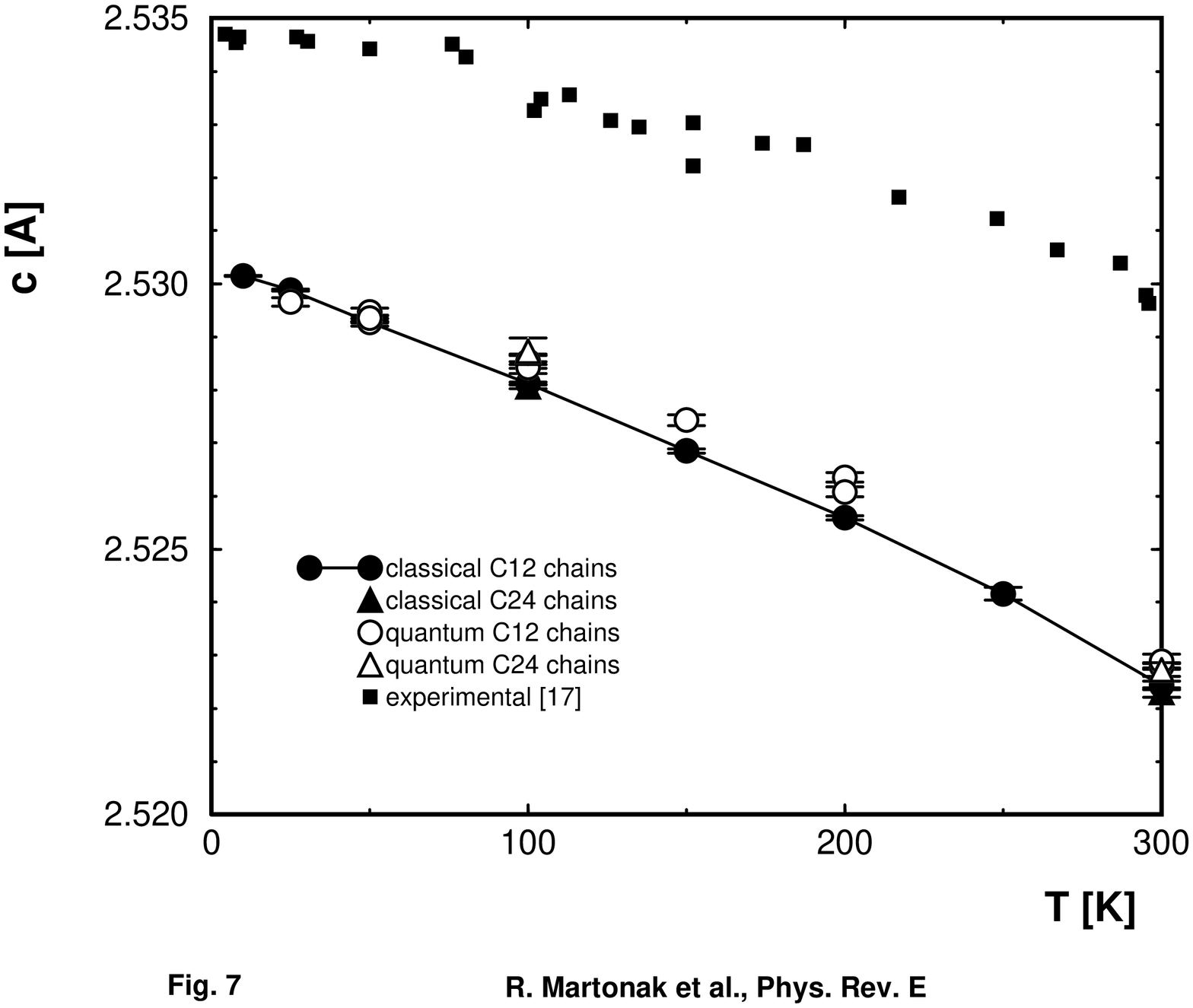,width=96mm,height=80mm,angle=-90}}
\end{picture}
\vskip 4.cm

\begin{picture}(96,80)
\put(0,0){\psfig{figure=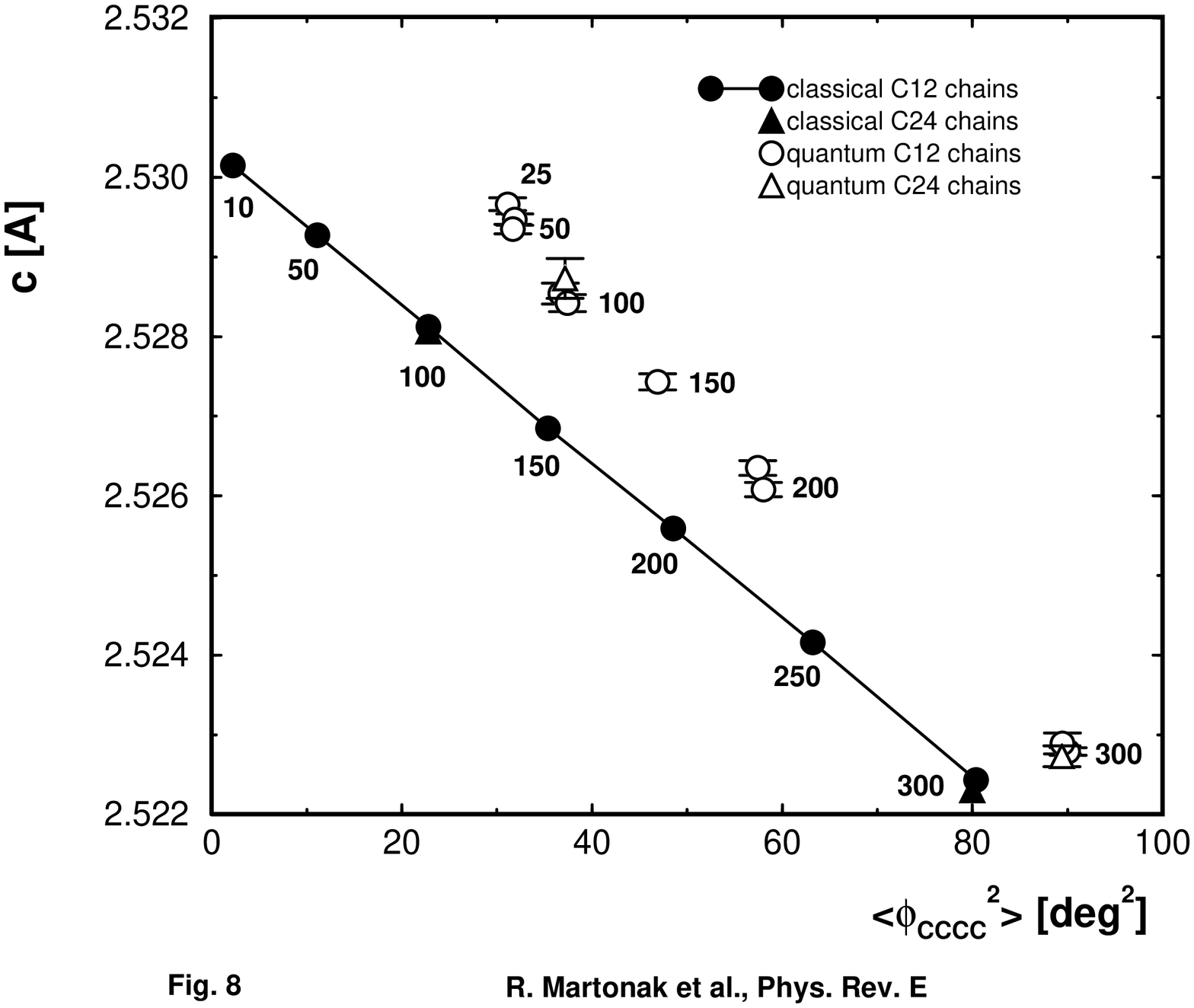,width=96mm,height=80mm,angle=-90}}
\end{picture}
\vfill\eject

\begin{picture}(96,80)
\put(0,0){\psfig{figure=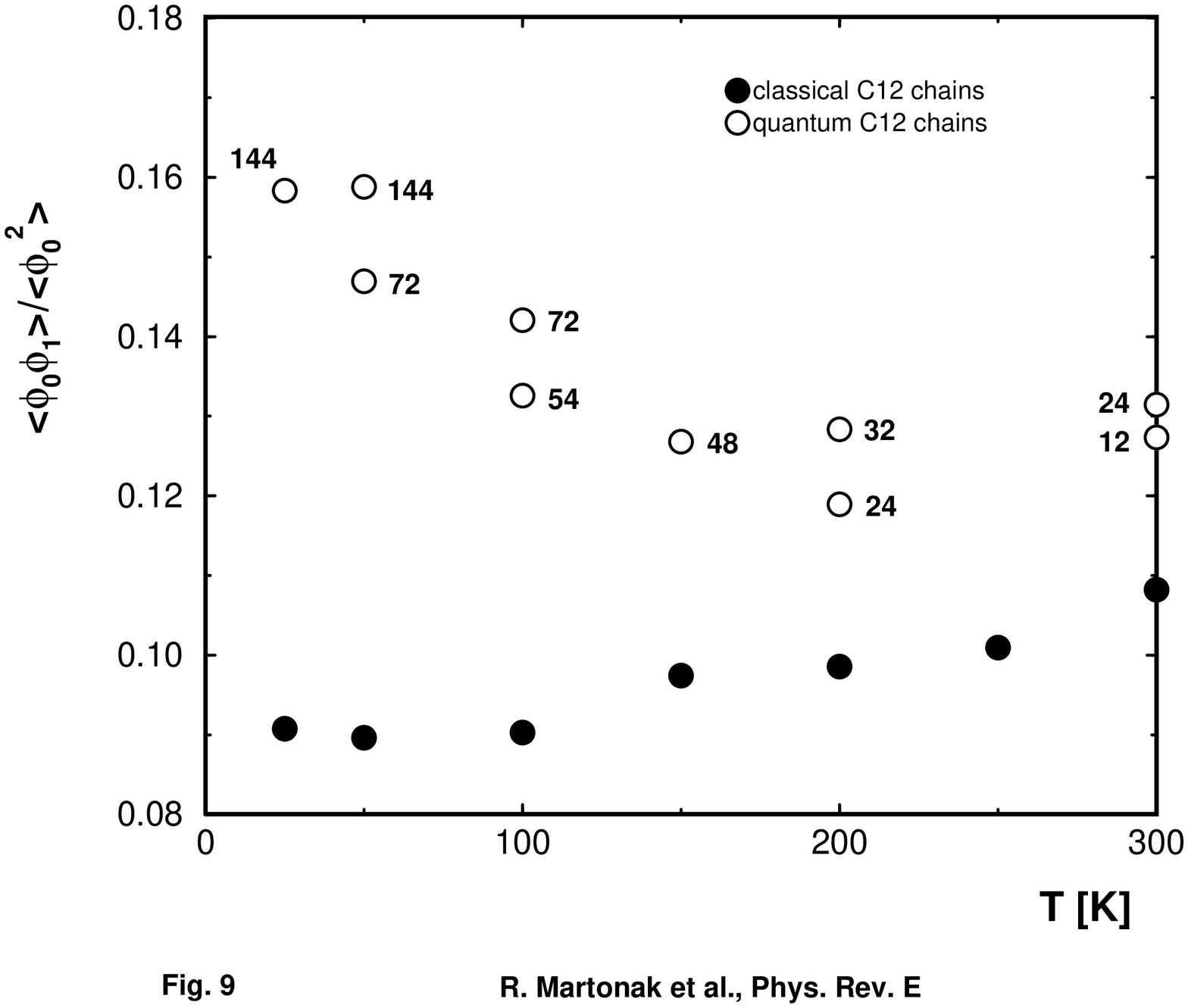,width=96mm,height=80mm,angle=-90}}
\end{picture}
\vskip 4.cm

\begin{picture}(96,80)
\put(0,0){\psfig{figure=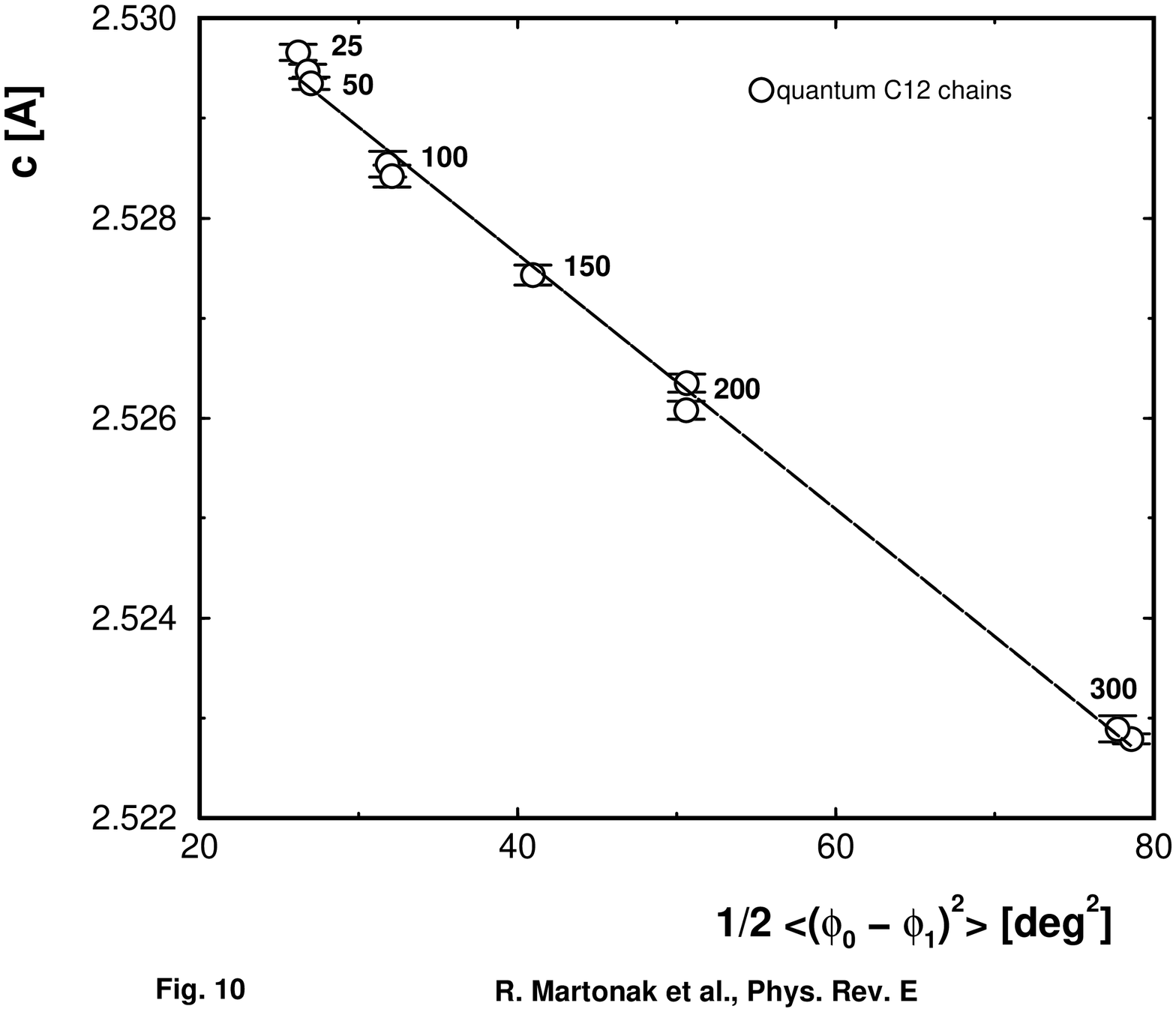,width=96mm,height=80mm,angle=-90}}
\end{picture}
\vfill\eject

\begin{picture}(96,80)
\put(0,0){\psfig{figure=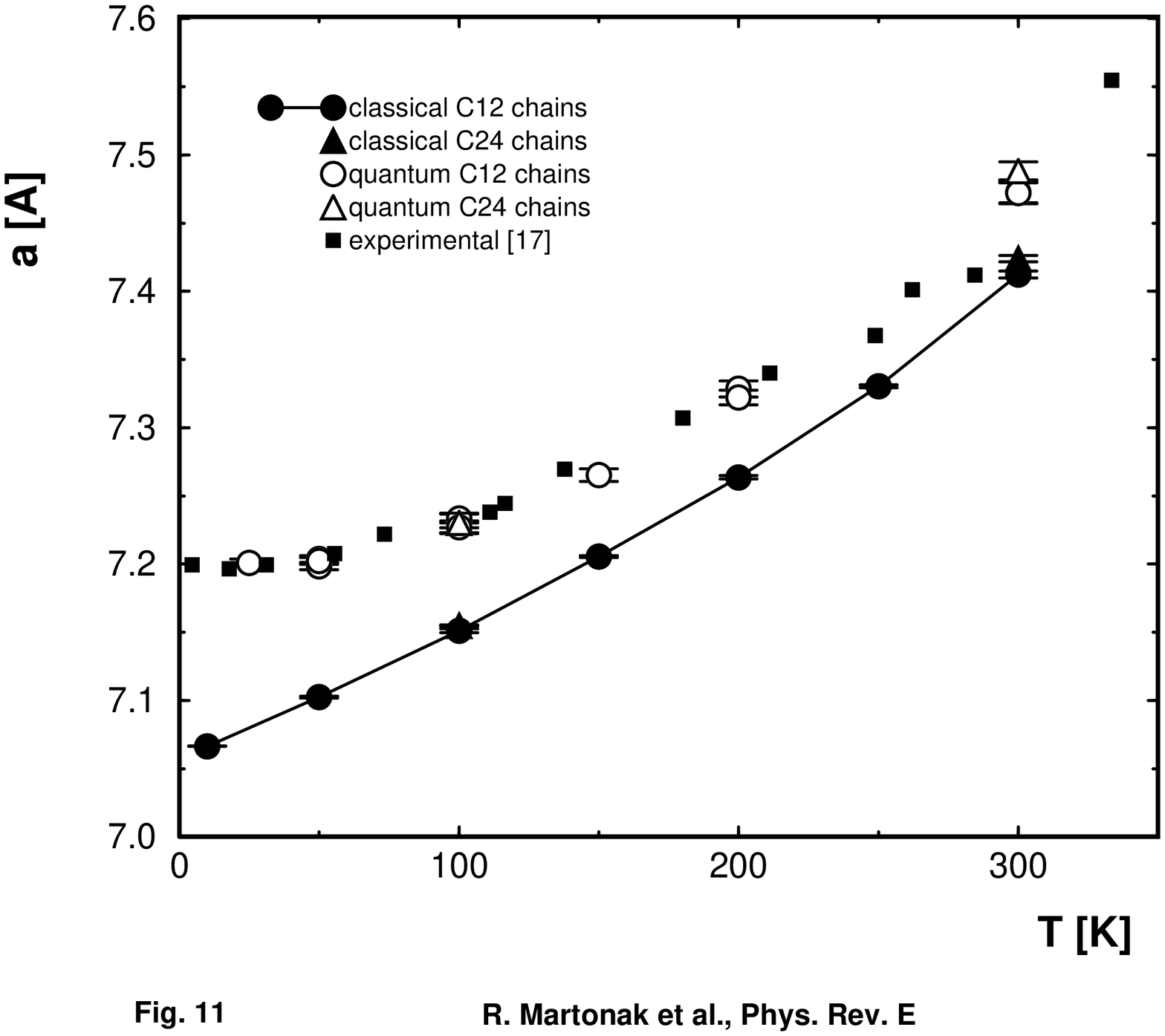,width=96mm,height=80mm,angle=-90}}
\end{picture}
\vskip 4.cm

\begin{picture}(96,80)
\put(0,0){\psfig{figure=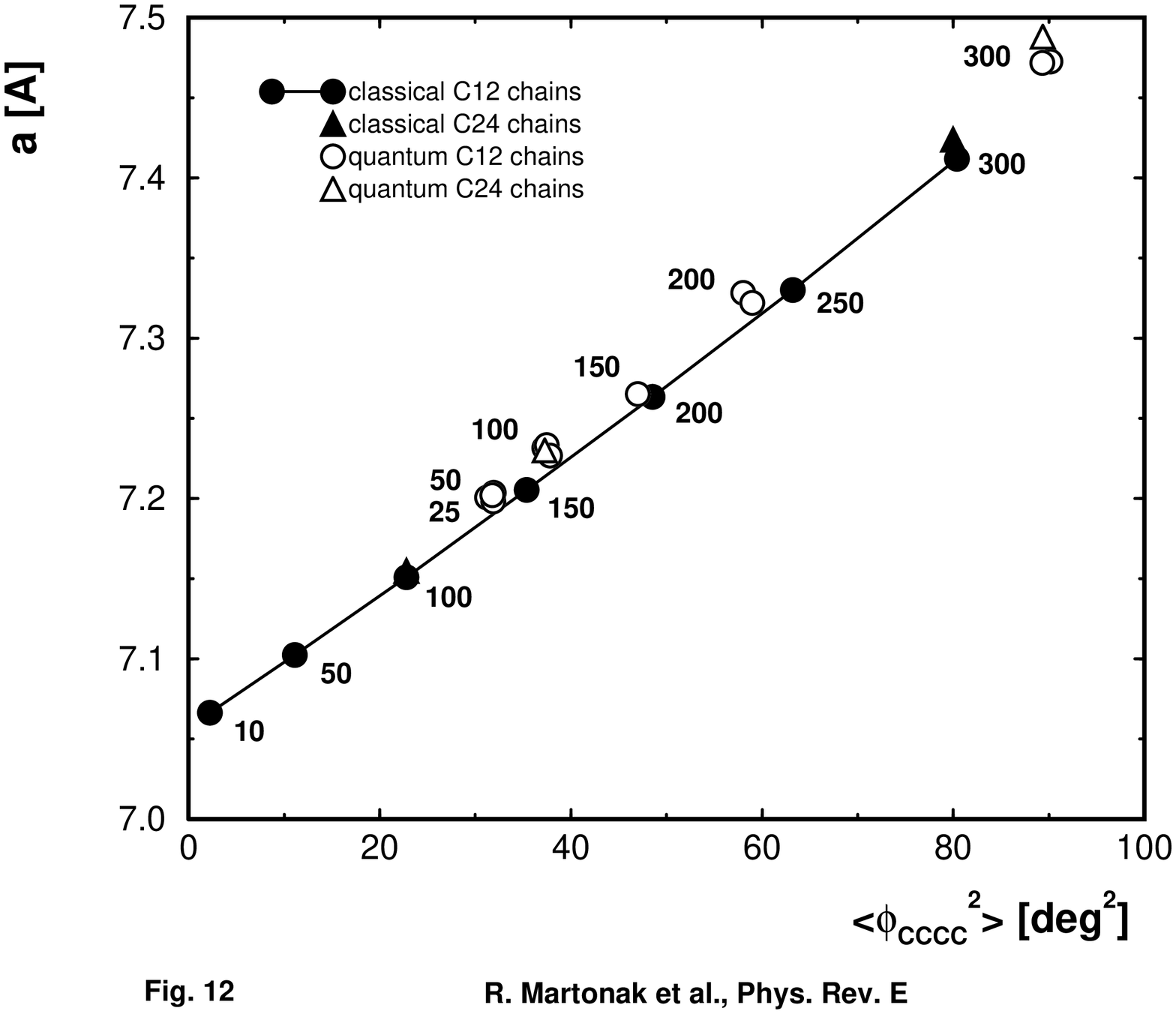,width=96mm,height=80mm,angle=-90}}
\end{picture}
\vfill\eject

\begin{picture}(96,80)
\put(0,0){\psfig{figure=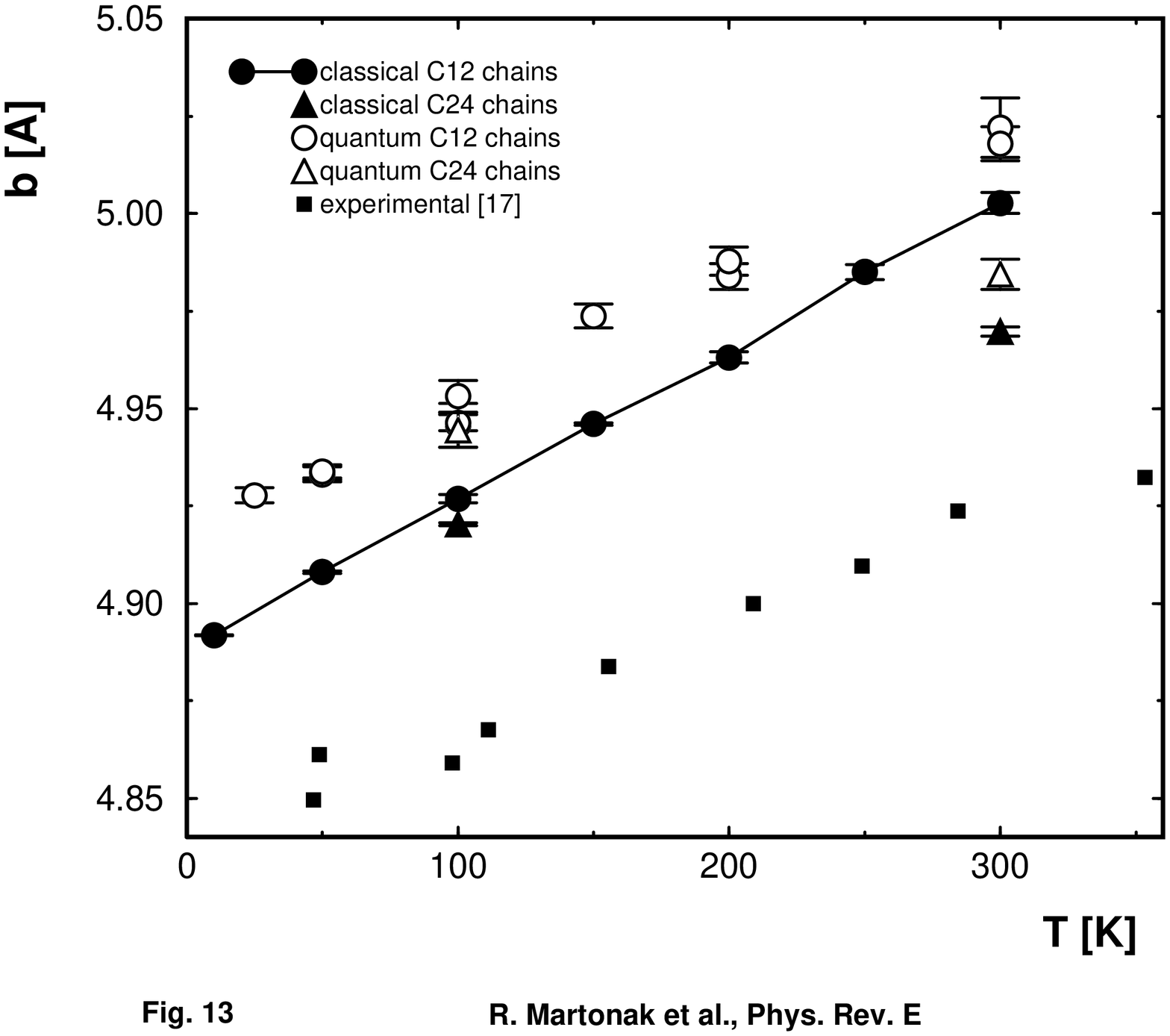,width=96mm,height=80mm,angle=-90}}
\end{picture}
\vskip 4.cm

\begin{picture}(96,80)
\put(0,0){\psfig{figure=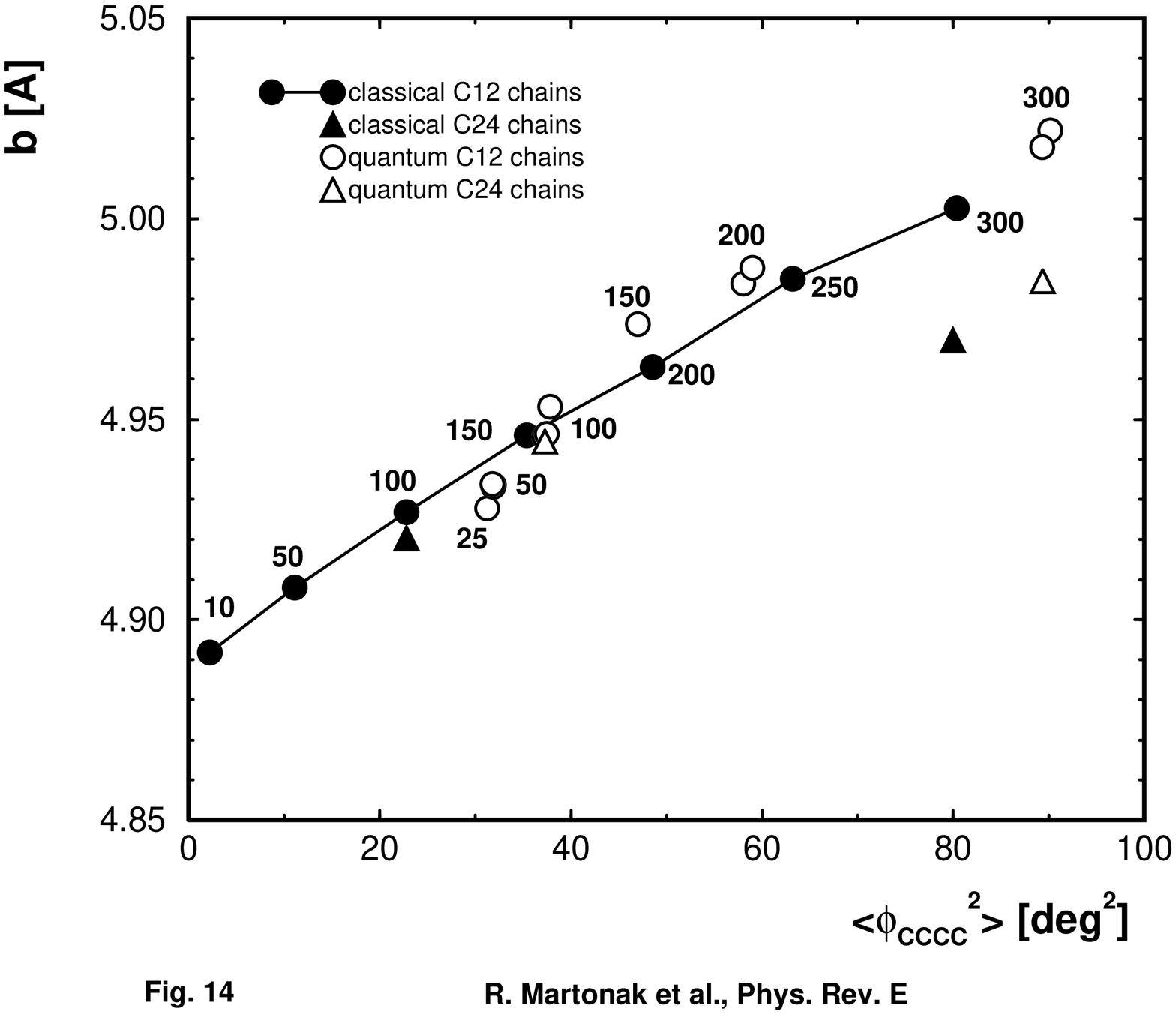,width=96mm,height=80mm,angle=-90}}
\end{picture}
\vfill\eject

\begin{picture}(96,80)
\put(0,0){\psfig{figure=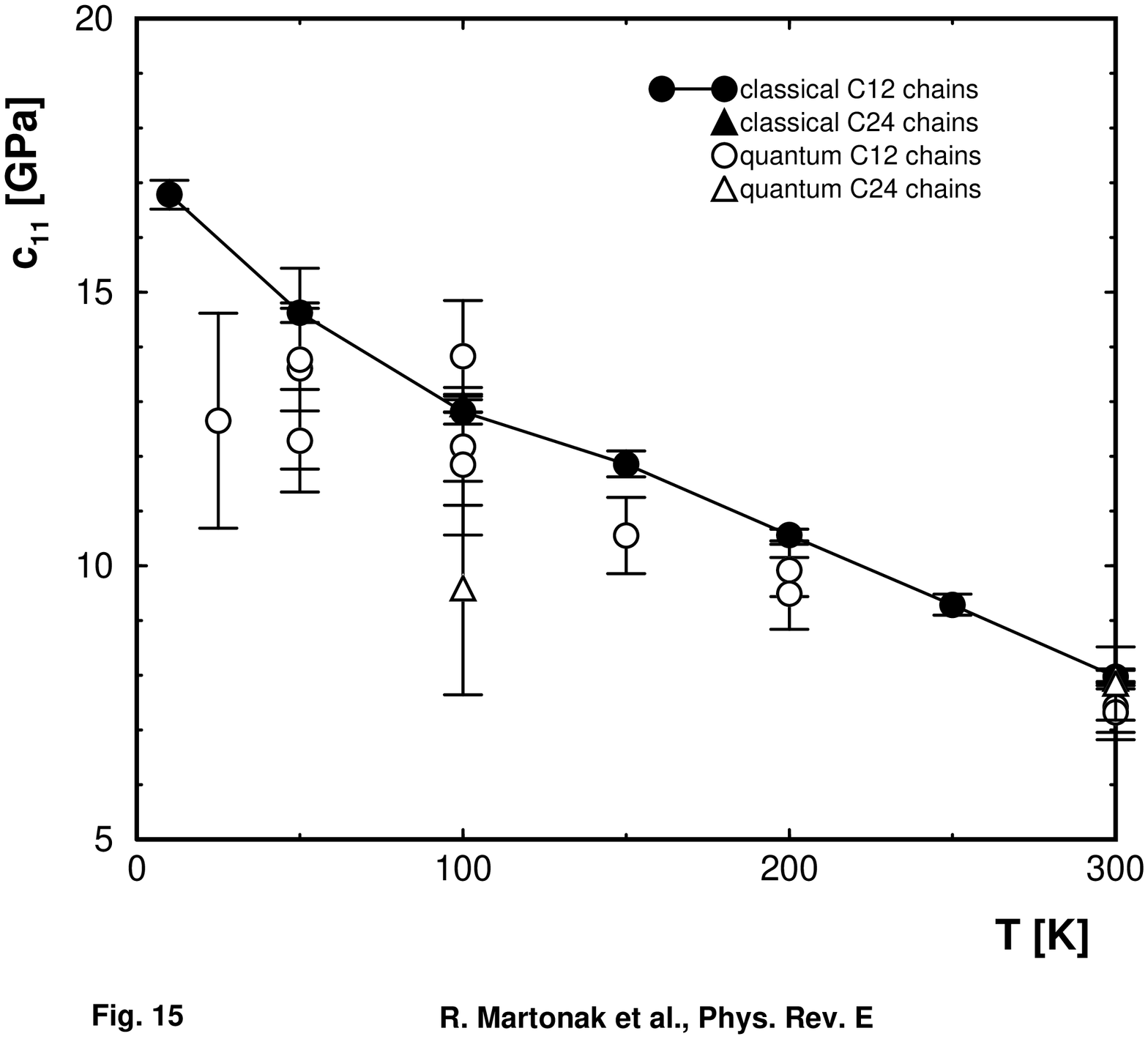,width=96mm,height=80mm,angle=-90}}
\end{picture}
\vskip 4.cm

\begin{picture}(96,80)
\put(0,0){\psfig{figure=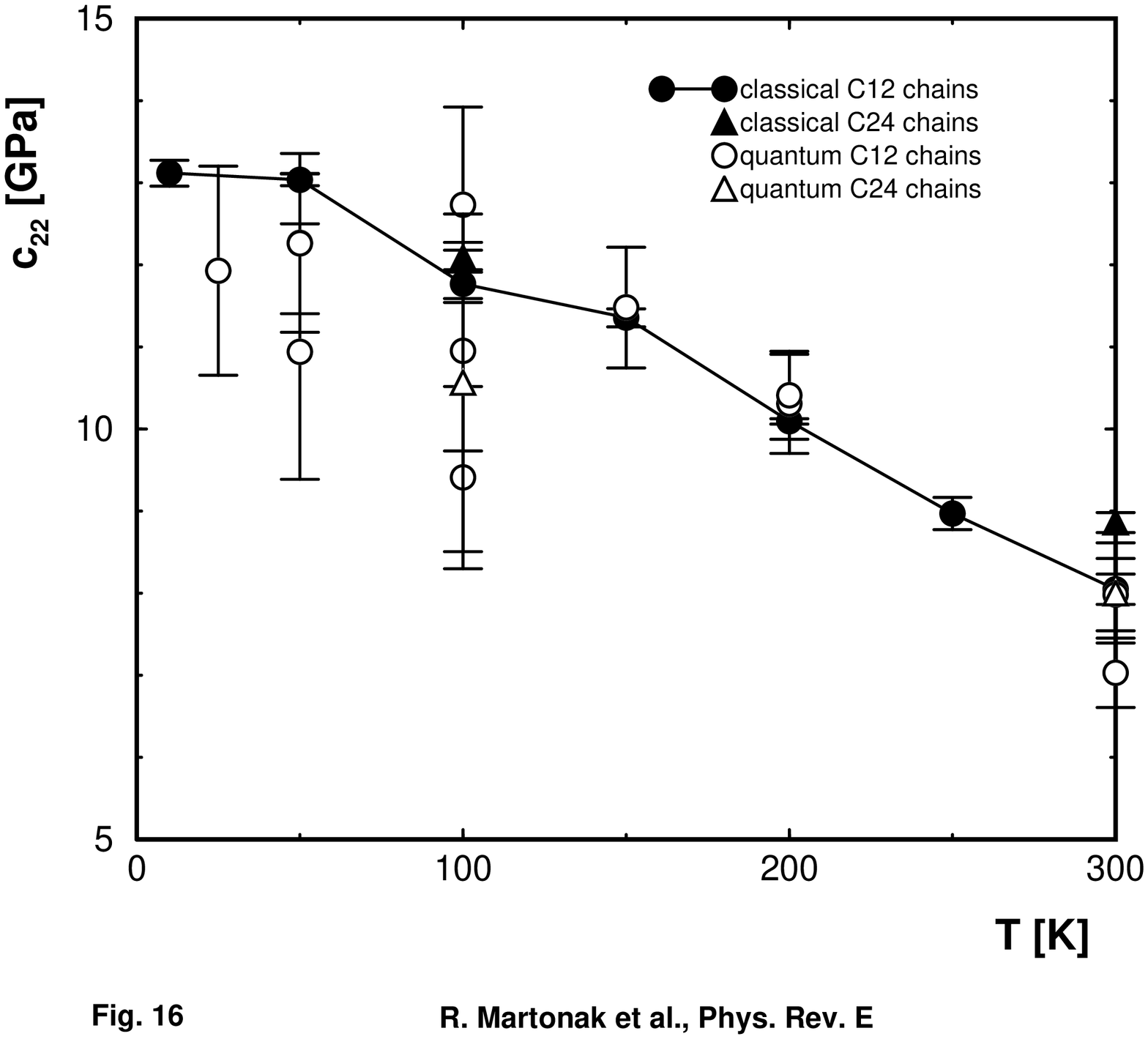,width=96mm,height=80mm,angle=-90}}
\end{picture}
\vfill\eject

\begin{picture}(96,80)
\put(0,0){\psfig{figure=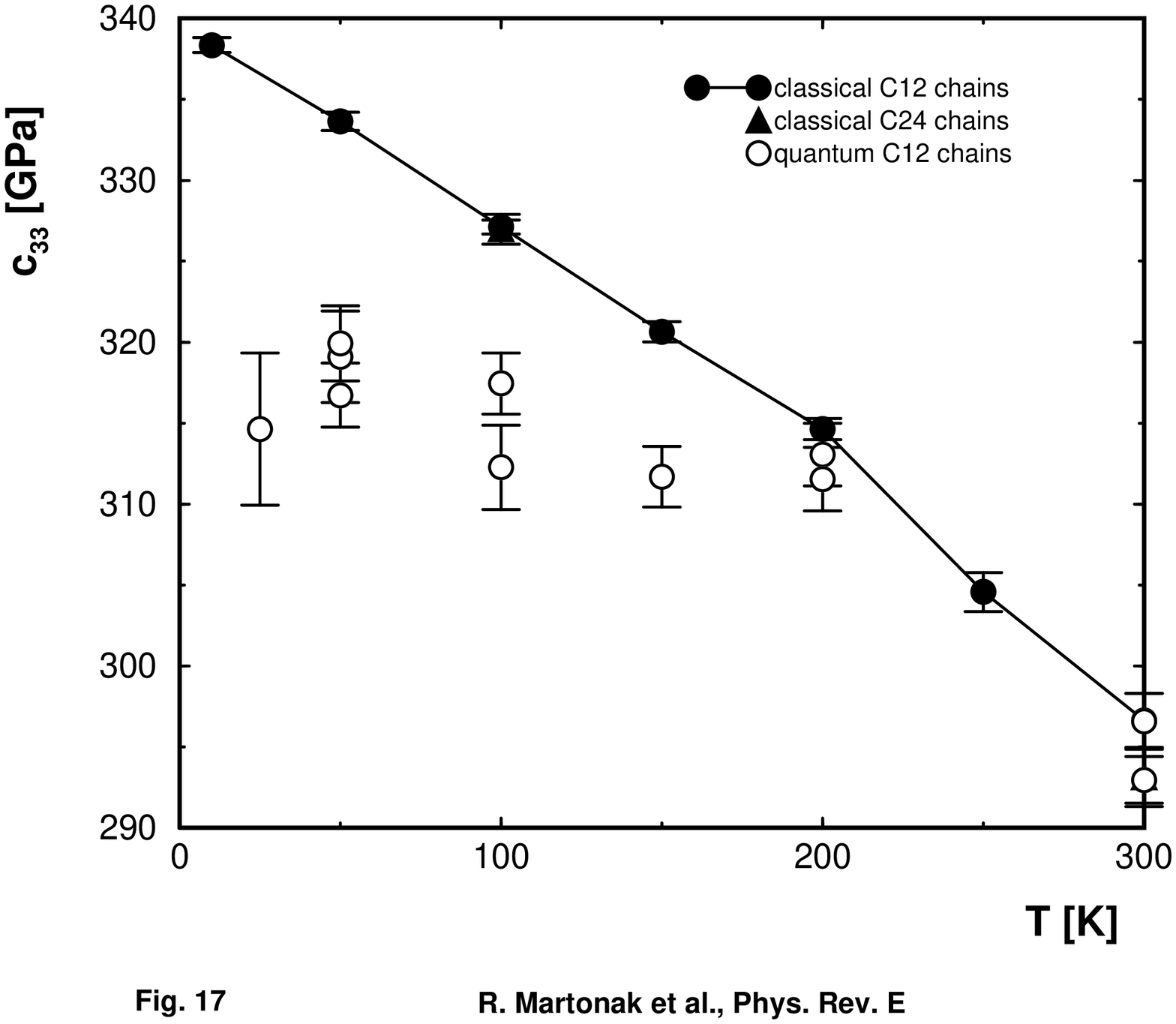,width=96mm,height=80mm,angle=-90}}
\end{picture}
\vskip 4.cm

\begin{picture}(96,80)
\put(0,0){\psfig{figure=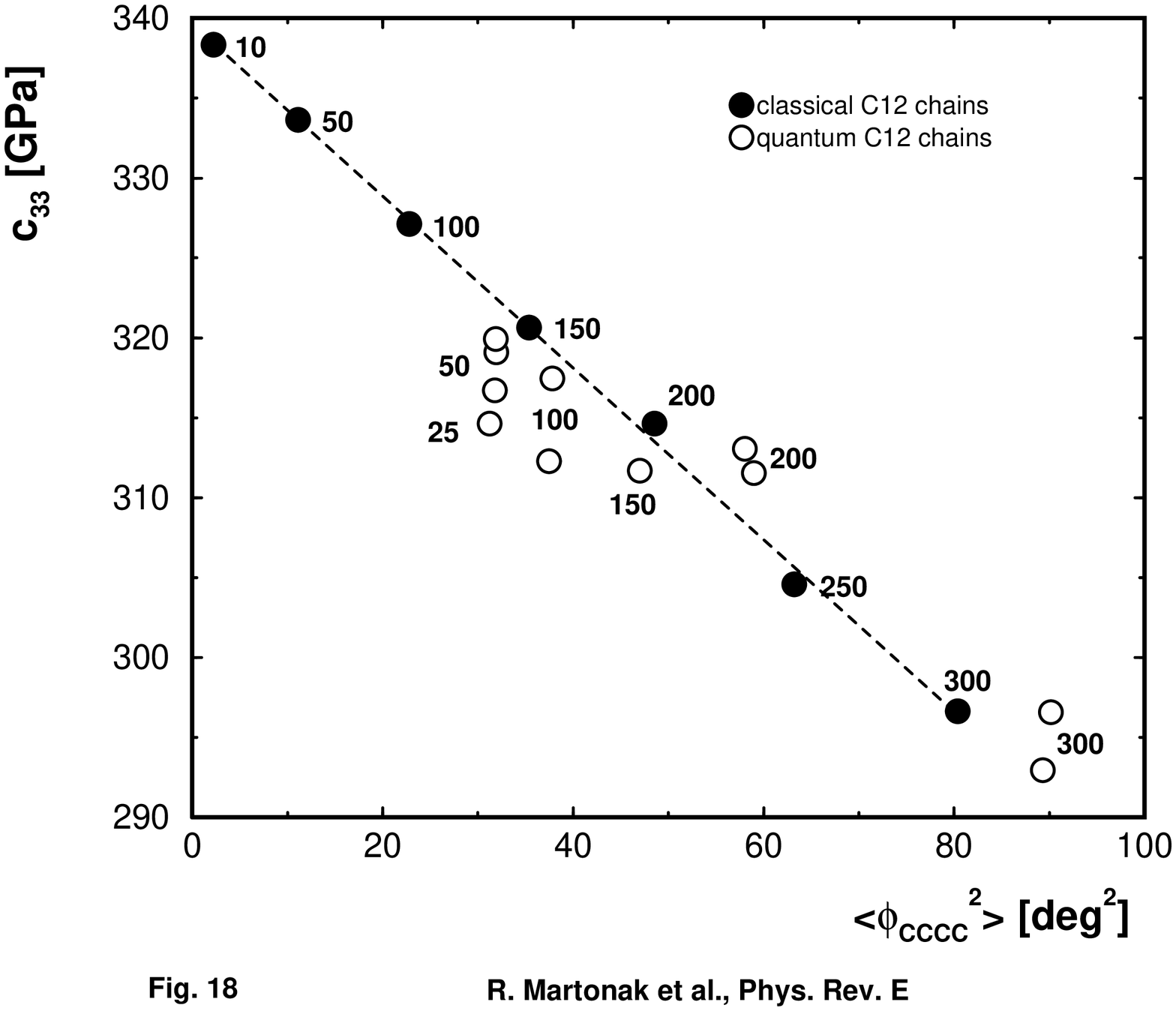,width=96mm,height=80mm,angle=-90}}
\end{picture}
\vfill\eject

\end{document}